\documentclass[twocolumn,showpacs,preprintnumbers,amsmath,amssymb,floatfix,prd]{revtex4}

\usepackage{citesort}
\usepackage{epsfig}

\begin{document}


\title{Global Analysis of Fragmentation Functions for Pions and Kaons\\
and Their Uncertainties}
\author{Daniel de Florian}\email{deflo@df.uba.ar}
\author{Rodolfo Sassot}\email{sassot@df.uba.ar}
\affiliation{Departamento de Fisica, Universidad de Buenos Aires, Ciudad Universitaria, Pab.\ 1 (1428)
Buenos Aires, Argentina}
\author{Marco Stratmann}\email{marco@ribf.riken.jp}
\affiliation{Radiation Laboratory, RIKEN, 2-1 Hirosawa, Wako, Saitama 351-0198, Japan}

\begin{abstract}
We present new sets of pion and kaon fragmentation functions obtained in NLO
combined analyses of single-inclusive hadron production in
electron-positron annihilation, proton-proton collisions, and deep-inelastic 
lepton-proton scattering with either pions or kaons identified in the final state.
At variance with all previous fits, the present analyses take into 
account data where hadrons of different electrical charge are identified, 
which allow to discriminate quark from anti-quark fragmentation functions without the need of 
non trivial flavor symmetry assumptions.
The resulting sets are in good agreement with all data analyzed,
which cover a much wider kinematical range than in previous fits.
An extensive use of the Lagrange multiplier technique is made in order to
assess the uncertainties in the extraction of the fragmentation functions and
the synergy from the complementary data sets in our global analysis.
\end{abstract}

\pacs{13.87.Fh, 13.85.Ni, 12.38.Bx}

\maketitle

\section{Introduction and Motivation}

The hadronization process turns partons produced in hard-scattering
reactions into the physical, colorless, non perturbative hadronic
bound states towards any hard interaction ultimately develops.
Within the standard framework of leading-twist collinear QCD
\cite{ref:framework}, processes with an observed hadron in the 
final-state can be
described in terms of perturbative hard scattering cross sections and
certain non-perturbative but universal functions: parton distributions, 
accounting for the partonic structure of the hadrons in the initial state
just before the interaction, and fragmentation functions,
encoding the details of the subsequent hadronization process
\cite{ref:collins-soper}.

These three ingredients are therefore the pillars of the perturbative
QCD (pQCD) description of hard interactions; their precise knowledge
has been crucial for its success in the past, and is imperative for the
ongoing and future high energy research programs \cite{Campbell:2006wx}.
In the last few
years, the improvement in each of these key areas has been remarkable. Higher order
QCD calculations have been explored and validated for most processes
up to next-to-leading order (NLO) accuracy, and are currently being extended
even beyond that point for some observables \cite{Vogt:2006bt}.

The knowledge on parton distributions has become increasingly precise
as a result of two decades of a wide variety of high precision measurements,
and strenuous efforts to update and enlarge periodically the corresponding QCD
analyses \cite{Thorne:2006zu,Pumplin:2005rh}. State-of-the-art sets of parton densities
agree with each other well within the already fairly 
small estimated uncertainties, and provide a picture
of the proton structure fully consistent with the data. For most observables,
the differences arising from the use of one or another modern set of parton
densities are negligibly small compared with the uncertainties in their
measurement or from unknown higher order corrections.

Also fragmentation functions have been rapidly evolving, following the path
of parton densities, however, without attaining yet the precision of the
latter \cite{ref:kretzer,ref:kkp,ref:akk,ref:hirai}.
Most of the information used to determine these distributions comes
essentially from electron-positron annihilation into charged hadrons.
These data have the important advantages of being very precise, thanks to
high statistics measurements from CERN-LEP and at SLAC
\cite{ref:alephdata,ref:delphidata,ref:opaldata,ref:opaleta,ref:tpcdata,ref:slddata} 
and clean, in the sense that the corresponding cross sections have no dependence on parton
densities.

In spite of these advantages, electron-positron annihilation data suffer
from many shortcomings. In the first place, the data give per se no information
on how to disentangle quark from anti-quark fragmentation as they always
refer to the charge sum for a certain hadron species, e.g., 
$\pi^++\pi^-$.
The information on how the individual quark flavors
fragment into hadrons depends crucially on so-called ``tagging''
techniques and the underlying assumptions implemented in the Monte-Carlo generators
employed \cite{ref:opaleta}. In addition to that,
at the mass scale of the $Z$-boson, i.e., for the bulk
of the electron-positron annihilation data, all electro-weak couplings become 
roughly equal and thus only flavor singlet combinations 
of fragmentation functions can be determined.

Also the gluon fragmentation is not exceedingly well constrained, since
the sub-leading NLO corrections for electron-positron annihilation
are too weak to determine it; quark-gluon mixing in the scale 
evolution of fragmentation functions is not enough of a constraint either, 
due to the lack of precise enough data at energy scales away from the $Z$-resonance. 
The data also become considerably less accurate and sparse at
large hadron energy fractions, leading to large uncertainties in that region.

Fortunately, in the last few years several one-particle inclusive
measurements coming from both proton-proton collisions and deep-inelastic 
lepton-nucleon scattering have matured enough as to
yield complementary information on the fragmentation process with
competing precision
\cite{ref:hermessidis,ref:phenixpion,ref:starpion,ref:brahms,ref:starkaon}.
These measurements not only probe fragmentation
in a complementary energy regime, but also weigh quite differently
contributions of individual parton flavors in the hadronization process.

Including these data in the extraction of fragmentation functions, 
not only increases statistics but, most importantly, yields a much more 
complete picture of the fragmentation process. For example, the complementary information
allows to relax and test certain rather stringent assumptions on the flavor symmetry
of fragmentation functions usually made. One can also make an independent check of the 
flavor tagging techniques implemented in electron-positron annihilation, and,
ultimately, scrutinize the fundamental ideas of QCD factorization 
and the universality of the fragmentation functions.

In this paper, we perform for the first time such a more comprehensive 
{\em global QCD analysis} to obtain new sets of pion and kaon fragmentation functions 
in agreement with the wealth of electron-positron annihilation data 
{\em and} also electron-proton and proton-proton observables.
The latter were either impossible to calculate with
previous sets or just not reproduced by them. 
As it should be expected, our new sets agree with previous
extractions \cite{ref:kretzer,ref:kkp,ref:akk,ref:hirai} in aspects
of fragmentation functions which are actually determined well by electron-positron data 
but show significant differences otherwise. 

In order to asses the uncertainties in the resulting fragmentation functions,
associated with both the uncertainties in the data and the theoretical estimate
of the observables, we have made an extensive use of the Lagrange multiplier
method \cite{ref:Stump:2001gu}. We study in detail the profiles of the $\chi^2$-function that
quantifies the agreement between the set of fragmentation functions,
more specifically their moments, and the data. This procedure gives a clear picture of the relative
uncertainty for each quark flavor and the gluon, which are found to be about 3\%
and 10\% for ``favored'' (valence) fragmentation functions in pions and kaons, respectively,
and of the order of 10\% and 20\% for the respective ``unfavored'' (sea) fragmentation
functions. The method allows also to assess the role and interplay of the
various data sets in constraining the different fragmentation
functions, illustrating the synergy characteristic of a global analysis.

In the following Section, we briefly summarize the QCD framework for
fragmentation functions and the different single-inclusive hadron
production processes included in the global analysis. In Sec.~III,
we outline the details of the analysis, discussing our choice for the
functional form used to parametrize the fragmentation functions 
at the initial scale of the scale evolution, and the
data sets included in the fit. We also outline
the implementation of the 
Mellin transform method
for a fast evaluation of the NLO cross sections, and the Lagrange multiplier
technique for assessing uncertainties. In Sec.~IV we discuss 
in great detail our results for both pion and for kaon fragmentation functions
and their uncertainties. We briefly summarize our results in Sec.~V.

\section{\label{sec:framework}QCD Framework for Fragmentation Functions and
         Single-Inclusive Hadron Production}
%
The pQCD framework for single-inclusive hadron production
in electron-positron annihilation, lepton-nucleon deep-inelastic
scattering, and hadron-hadron collisions is in place for
quite some time now, and calculations at NLO accuracy
are ``state-of-the-art'' throughout.
In each case one exploits the factorization theorem \cite{ref:framework},
which states that the cross section can be decomposed into appropriate convolutions of
perturbatively calculable partonic hard scattering cross sections and certain
combinations of non-perturbative parton distribution and
fragmentation functions.
We can restrict ourselves to a brief summary of
theoretical framework relevant for our global QCD analysis of
fragmentation functions to set up our notation in which we closely
follow Refs. \cite{ref:dsv-lambda} and \cite{ref:jsv-pion}.
%
\subsection{\label{sec:ffs} Properties of Fragmentation Functions}
%
A field theoretical definition of fragmentation functions $D_f^H$ in terms of 
bi-local operators was given in Ref.~\cite{ref:collins-soper} and
reads for quarks (up to kinematical pre-factors)
\begin{eqnarray}
\label{eq:operator}
D_q^H(z) \!\!&\propto& \!\!\! \int dx^- e^{-iP_H^+ x^-/z} \nonumber \\
\nonumber
&&\!\!\!\text{Tr} \Big[ \gamma^+ \langle 0|
\Psi(0) {\cal{P}} |H(P_H^+)X\rangle \langle H(P_H^+)X| {\cal{P^{\prime}}} \bar{\Psi}(x)
|0\rangle \Big]\\
\end{eqnarray}
and similarly for gluons. ${\cal{P}}$ and ${\cal{P^{\prime}}}$ denote the necessary gauge-links
to render (\ref{eq:operator}) gauge-invariant. The simple parton model interpretation of
$D_f^H(z)$ as the probability for a parton $f$ to produce a hadron $H$ with fraction $z$ of
its momentum is recovered in light-cone gauge where ${\cal{P}}={\cal{P^{\prime}}}=1$.
Since a specific hadron $H$ with light-cone momentum $P_H^+$ is observed in the final-state,
a local operator product expansion (OPE) does not apply. For fully inclusive parton densities,
the OPE is the basis for first principle computations of some of their integer moments
within ``lattice QCD''. Similar calculations cannot be pursued for fragmentation functions.

The scale dependence of the fragmentation functions $D_f^H$ 
is calculable in pQCD and
governed by renormalization group equations very similar to those for
parton densities. For instance, the singlet evolution
equation schematically reads
\begin{eqnarray}
\label{eq:singevol}
\frac{d}{d\ln Q^2} \vec{D}^H (z,Q^2) = \left[
\hat{P}^{(T)}\otimes \vec{D}^H\right](z,Q^2),
\end{eqnarray}
where
\begin{equation} \label{eq:singlet}
\vec{D}^H \equiv \left( \begin{array}{c} D_{\Sigma}^H \\ D_g^H \\
\end{array}\right),\,\,\,
D_{\Sigma}^H \equiv \sum_q (D_q^H+ D_{\bar{q}}^H)
\end{equation}
and
\begin{eqnarray}
\label{eq:pmatrix}
\renewcommand{\arraystretch}{1.3}
\hat{P}^{(T)} \equiv \left( \begin{array}{cc}
P_{qq}^{(T)} &  2n_f P_{gq}^{(T)} \\
\frac{1}{2n_f} P_{qg}^{(T)} & P_{gg}^{(T)} \\
\end{array}\right) \; .
\end{eqnarray}
is the matrix of the singlet {\em timelike} evolution kernels.
The NLO splitting functions $P_{ij}^{(T)}$ have been computed in
\cite{ref:nlo-kernels,ref:rijken} or can be related to the
corresponding spacelike kernels by proper
analytic continuation \cite{ref:sv-kernels}.

The range of applicability for fragmentation functions
as defined above is severely limited to medium-to-large values of $z$.
On the one hand, the timelike evolution kernels in (\ref{eq:pmatrix})
develop a strong singular behavior as $z\to 0$, and, on the other hand,
the produced hadrons are considered to be massless.
More specifically, the splitting functions $P_{gq}^{(T)}(z)$ and
$P_{gg}^{(T)}(z)$ have a dominant, large logarithmic piece $\simeq \ln^2 z/z$
in their NLO part, which ultimately leads to negative fragmentation functions
for $z\ll 1$ in the course of the $Q^2$ evolution and, perhaps, to
unphysical, negative cross sections, even if the evolution starts with
positive distributions at some scale $Q_0<Q$.
At small $z$, also finite mass corrections
proportional to $M_H/(s z^2)$ become more and more important.
While there are ways to resum the singular small-$z$ behavior to all
orders in $\alpha_s$, there is no systematic or unique way to correct for
finite hadron masses, for instance by introducing some ``re-scaled'' variable $z^{\prime}$
in SIA. Inseparably entwined with mass effects are other power corrections or
``dynamical higher twists''.

Anyway, including small-$z$ resummations or mass corrections in one way or the other
in the analysis of hadron production rates is not compatible with the
factorization theorem and the definition of fragmentation functions outlined
above. ``Resummed'' or ``mass corrected'' fragmentation functions should not be
used with fixed order expressions for, say, the semi-inclusive deep-inelastic
production of a hadron, $e N \rightarrow e' H X$, discussed in
Sec.~\ref{sec:sidis}.
Therefore we limit ourselves in our global analysis to kinematical
regions where mass corrections and the influence of the
singular small-$z$ behavior of the evolution kernels is negligible.
It turns out that a cut $z>z_{\min}=0.05\, (0.1)$ is sufficient for data on
pion (kaon) production.

Finally, conservation of the momentum of the fragmenting parton $f$ in
the hadronization process is summarized by a sum rule stating that
\begin{equation}
\label{eq:sumrule}
\sum_H \int_0^1 dz z D_i^H(z,Q^2) = 1,
\end{equation}
i.e., each parton will fragment with 100$\%$ probability into some hadron $H$.
Equation~(\ref{eq:sumrule}) is compatible with the evolution kernels
in the ${\overline{\mathrm{MS}}}$ scheme, although not for each individual
contribution $\int_0^1 dz z D_i^H(z,Q^2)$. Of course, the sum rule
(\ref{eq:sumrule}) should be dominated, perhaps almost saturated, by the
fragmentation into the lightest hadrons such as pions and kaons.
The unstable small-$z$ behavior, however, prevents Eq.~(\ref{eq:sumrule})
from being a viable constraint in a global analysis. Only truncated moments
$\int_{z_{\min}}^1 dz z D_i^H(z,Q^2)$ are meaningful.

\subsection{Single-inclusive $e^+e^-$ Annihilation}
%
The cross sections for the single-inclusive $e^+e^-$ annihilation (SIA) into
a specific hadron $H$,
\begin{equation}
e^+e^- \rightarrow (\gamma,\,Z) \rightarrow H,
\end{equation}
at a center-of-mass system (c.m.s.)~energy $\sqrt{s}$
and integrated over the production angle
can be written as \cite{ref:aempi,ref:fupe}
\begin{equation}
\label{eq:ee-xsec}
\frac{1}{\sigma_{tot}} \frac{d\sigma^H}{dz} = \frac{\sigma_0}
{\sum_q \hat{e}_q^2}
\left[ 2\, F_1^{H}(z,Q^2) +  F_L^{H}(z,Q^2) \right].
\end{equation}
The energy $E_H$ of the observed hadron scaled to the beam
energy $Q/2=\sqrt{s}/2$ is denoted by
$z\equiv 2 p_H\cdot q/Q^2 = 2 E_H/\sqrt{s}$ with
$Q$ being the momentum of the intermediate $\gamma$ or $Z$ boson.
\begin{equation}
\label{eq:sigmatot}
\sigma_{tot}=\sum_q \hat{e}_q^2\; \sigma_0
\left[1+\frac{\alpha_s(Q^2)}{\pi}\right]
\end{equation}
is the total cross section for $e^+e^- \rightarrow hadrons$ including its NLO
${\cal{O}}(\alpha_s)$ correction and $\sigma_0=4\pi\alpha^2(Q^2)/s$.
The sums in (\ref{eq:ee-xsec}) and (\ref{eq:sigmatot}) run over the $n_f$
active quark flavors $q$, and the $\hat{e}_q$ are the corresponding
appropriate electroweak charges (see App.~A of Ref.~\cite{ref:dsv-lambda}
for details).

To NLO accuracy, the unpolarized ``time-like'' structure functions
$F_1^{H}$ and $F_L^{H}$ in (\ref{eq:ee-xsec}) are given by
\begin{eqnarray}
\label{eq:f1nlo}
\nonumber
2 F_1^{H}(z,Q^2) &=& \sum_q \hat{e}_q^2\;
\Bigg\{ \left[ D_q^H (z,Q^2) + D_{\bar q}^H (z,Q^2) \right]\\
\nonumber
&& + \frac{\alpha_s(Q^2)}{2\pi} \left[ C_q^1 \otimes
(D_q^H+D_{\bar q}^H) \right.\\
&& \left. + C_g^1 \otimes D_g^H \right] (z,Q^2) \Bigg\}, \\
\nonumber
\label{eq:flnlo}
F_L^{H}(z,Q^2) &=& \frac{\alpha_s(Q^2)}{2\pi} \sum_q \hat{e}_q^2 \;
\left[ C_q^L \otimes (D_q^H+D_{\bar q}^H) \right. \\
&& \left. +C_g^L \otimes D_g^H
\right](z,Q^2),
\end{eqnarray}
with $\otimes$ denoting a standard convolution.
The relevant NLO coefficient functions $C_{q,g}^{1,L}$
in the $\overline{\mathrm{MS}}$ scheme
can be found in App.~A of Ref.~\cite{ref:dsv-lambda}.

We note that the longitudinal structure function $F_L$ in Eq.(\ref{eq:flnlo})
receives its leading nonzero (finite and scheme independent) contribution
at ${\cal{O}}(\alpha_s)$. We treat, however, the ${\cal{O}}(\alpha_s)$
expressions in (\ref{eq:flnlo}) as sub-leading (=NLO) in calculations
of the {\em total} (longitudinal plus transverse) cross section (\ref{eq:ee-xsec}).
For predictions of only the longitudinal cross section at NLO, the
${\cal{O}}(\alpha_s^2)$ corrections \cite{ref:rijken} should be included.
However, such measurements are not available for identified pions or kaons considered
in this analysis.

\subsection{\label{sec:sidis} Semi-Inclusive Deep-Inelastic Scattering}
%
The cross section for the semi-inclusive deep-inelastic
production of a hadron, $e N \rightarrow e' H X$, is proportional
to certain combinations of both the parton distributions of
the nucleon $N$ and the fragmentation functions for the hadron $H$.
It can be written in factorized form in a way very similar to the
fully inclusive DIS case \cite{ref:aempi,ref:fupe,ref:graudenz,ref:dsv-lambda}:
\begin{eqnarray}
\label{eq:sidis-xsec}
\nonumber
\frac{d\sigma^H}{dx\, dy\, dz_H} &=&
\frac{2\, \pi\alpha^2}{Q^2}
\left[ \frac{(1+(1-y)^2)}{y} 2\, F_1^{H}(x,z_H,Q^2) \right. \\
&& \left. +
\frac{2 (1-y)}{y} F_L^{H}(x,z_H,Q^2) \right],
\end{eqnarray}
with $x$ and $y$ denoting the usual DIS scaling variables $(Q^2=s x y)$, and
where \cite{ref:aempi,ref:fupe} $z_H\equiv p_H\cdot p_N/p_N\cdot q$
with an obvious notation of the four-momenta, and with $-q^2\equiv Q^2$.
Strictly speaking, Eq.\ (\ref{eq:sidis-xsec}) and the variable $z_H$ only apply
to hadron production in the current fragmentation region. This is usually
ensured by a cut $x_F>0$ on the Feynman-variable representing the fractional
longitudinal c.m.s.\ momentum. If necessary, target fragmentation could be accounted
for by transforming to the variable \cite{ref:targetfrag,ref:graudenz}
$z_H \rightarrow z \equiv \frac{E_H}{E_N (1-x)}$,
the energies $E_H$, $E_N$ defined in the c.m.s.\ frame of the nucleon and the
virtual photon, and by introducing the so-called ``fracture functions''
\cite{ref:targetfrag}.

The structure functions $F_1^{H}$ and $F_L^{H}$ in (\ref{eq:sidis-xsec})
are given at NLO by
\begin{eqnarray}
\label{eq:f1sidis}
\nonumber
2 F_{1}^{H}(x,z_H,Q^2) &\!\!=\!\!& \sum_{q,\overline{q}} e_q^2 \Bigg\{  q (x,Q^2)  D^H_q (z_H,Q^2) \\
\nonumber
&& +\frac{\alpha_s(Q^2)}{2\pi} \bigg[ q
\otimes  C^1_{qq} \otimes D^H_q\\
\nonumber
&& + q  \otimes  C^1_{gq} \otimes D^H_g  \\
&& + g  \otimes
C^1_{qg} \otimes D^H_q \bigg] (x,z_H,Q^2)\! \Bigg\}, \\
\label{eq:flsidis}
\nonumber
F_{L}^{H}(x,z_H,Q^2) &\!\!=\!\!& \frac{\alpha_s(Q^2)}{2\pi}
\sum_{q,\overline{q}} e_q^2  \bigg[ q
\otimes  C^L_{qq} \otimes D^H_q   \\
\nonumber
&& +q  \otimes  C^L_{gq} \otimes D^H_g \\
&& + g  \otimes
C^L_{qg} \otimes D^H_q \bigg] (x,z_H,Q^2),
\end {eqnarray}
with the NLO ($\overline{\mathrm{MS}}$) coefficient functions
$C^{1,L}_{ij}$ \cite{ref:aempi,ref:fupe,ref:graudenz,ref:dsv-lambda}.

In our global analysis of fragmentation functions we will make use of
(preliminary) data for charged pion and charged kaon multiplicities taken
by the HERMES experiment \cite{ref:hermessidis}. The multiplicities $(1/N_{\text{DIS}}) dN^H/dzdQ^2$
are defined as the ratio of the semi-inclusive deep-inelastic scattering (SIDIS) 
cross section (\ref{eq:sidis-xsec}) in a certain bin of, say, $Q^2$ and $z$, to the totally inclusive DIS rate.
The particular value of this data in the global analysis emerges from
the sensitivity to individual quark and anti-quark flavors in the fragmentation
process which is not accessible from $e^+e^-$ annihilation.

\subsection{Hadron-Hadron Collisions}
The single-inclusive production of a hadron $H$ at
high transverse momentum $p_T$ in hadron-hadron collisions is
also amenable to QCD perturbation theory.
Up to corrections suppressed by inverse powers of $p_T$,
the differential cross section can be written in factorized form
as \cite{ref:aversa,ref:jsv-pion}
\begin{equation}
\label{eq:pp-xsec}
E_H \frac{d^3 \sigma}{dp_H^3}  = \sum_{a,b,c} f_a \otimes f_b  \otimes d\hat{\sigma}_{ab}^{c}
\otimes D_c^{H},
\end{equation}
where the sum is over all contributing partonic channels $a+b\to
c + X$, with $d\hat{\sigma}_{ab}^{c}$ the associated partonic cross
section. $d\hat{\sigma}_{ab}^{c}$ can be expanded as a power series
in the strong coupling $\alpha_s$ and the ${\cal{O}}(\alpha_s^3)$
NLO corrections are available \cite{ref:aversa,ref:jsv-pion}.
As always, the factorized structure (\ref{eq:pp-xsec}) forces one to
introduce into the calculation scales of the order of the hard scale
in the reaction -- but not specified further by the theory -- that separate the
short- and long-distance contributions. We have suppressed the
explicit dependence on these renormalization and factorization scales
in Eq.~(\ref{eq:pp-xsec}), for details, see, e.g., Ref.~\cite{ref:jsv-pion}.

In studies and quantitative analyzes of hadronic cross sections,
NLO corrections are of particular importance and generally
indispensable in order to arrive at a firm theoretical
prediction for (\ref{eq:pp-xsec}). Since NLO corrections are known
to be significant, LO approximations usually significantly undershoot the
available data. In addition, hadronic reactions suffer from much
enhanced theoretical uncertainties than the reactions described above
due to the presence of more non-perturbative, scale dependent
functions. The dependence on the unphysical factorization and
renormalization scales can be only controlled and quantified at NLO
(or beyond).

As will be discussed below, the special value of hadronic cross sections in a global analysis
of fragmentation functions is their enhanced sensitivity to the gluon
fragmentation function through the dominance of $gg\to gX$ processes
for hadrons produced at low-to-medium transverse momenta and
their sensitivity to fragmentation at very high $z$.
Charge separated data for $H=\pi^{\pm}$ and $K^{\pm}$ provide additional information on the
flavor separation of the $D_i^H$.



\section{\label{sec:analysis-outline}Outline of the Analysis}
%
In this Section, we outline the details of our analysis. More specifically,
we discuss our choice of parametrization, the selection of data sets,
treatment of experimental normalization uncertainties, and
how we determine the parameters by means of a 
global $\chi^2$ minimization.
We also briefly sketch how we make use of Mellin moments to include exact
NLO expressions for the cross sections (\ref{eq:ee-xsec}),
(\ref{eq:sidis-xsec}), and (\ref{eq:pp-xsec}) in our analysis and
how we assess uncertainties in the extraction of fragmentation functions
with the help of the Lagrange multiplier technique.
%
\subsection{Parametrization}
%
All recent analyses of fragmentation functions are based exclusively on
SIA data \cite{ref:kretzer,ref:kkp,ref:akk,ref:hirai} and
have chosen the most simple functional form  $N_i z^{\alpha_i} (1-z)^{\beta_i}$
to parametrize the $D_i^H$ at some initial scale $\mu_0$ for the $Q^2$-evolution
(\ref{eq:singevol}).
The structure of the SIA cross section (\ref{eq:ee-xsec})-(\ref{eq:flnlo})
allows to extract only information on $D_{q+\bar{q}}^{\pi^+ + \pi^-}$ from data
(similarly for kaons). Without assumptions it is impossible to distinguish
``favored'' or ``valence'' from
``unfavored'' or ``sea'' fragmentation, for instance, $D_u^{\pi^+}$ from $D_{\bar{u}}^{\pi^+}$
where $|\pi^+\rangle =| u\bar{d}\rangle$.
This is a serious limitation of all present analyses \cite{ref:kretzer,ref:kkp,ref:akk,ref:hirai},
as the obtained fragmentation functions cannot be used to compare
to a wealth of recent data on the production of charged pions and kaons
in SIDIS \cite{ref:hermessidis} or proton-proton collisions \cite{ref:brahms}.
In Ref.~\cite{ref:kretzer} a linear suppression factor
$D_{\bar{u}}^{\pi^+}/D_{u}^{\pi^+}=(1-z)$ was {\em assumed} to break this
``deadlock''. This was later shown to be in fair agreement with charged pion multiplicities
in SIDIS from HERMES \cite{ref:hermessidis} within a LO combined analysis of
SIA and SIDIS data \cite{ref:kretzer2}; see also Fig.~\ref{fig:sidis-pion}
and discussions below.

In our global analysis we will determine for the first time individual fragmentation
functions for quark and anti-quarks for all flavors as well as gluons from data.
To accommodate also the experimental information from lepton-nucleon and
hadron-hadron scattering data, we adopt a somewhat more flexible input distribution
than in \cite{ref:kretzer,ref:kkp,ref:akk,ref:hirai}
\begin{equation}
\label{eq:ff-input}
D_i^H(z,\mu_0) =
\frac{N_i z^{\alpha_i}(1-z)^{\beta_i} [1+\gamma_i (1-z)^{\delta_i}] }
{B[2+\alpha_i,\beta_i+1]+\gamma_i B[2+\alpha_i,\beta_i+\delta_i+1]},
\end{equation}
where $B[a,b]$ represents the Euler Beta-function and $N_i$ is
normalized such to represent the contribution of $D_i^H$ to the sum rule (\ref{eq:sumrule}).
A more restrictive initial parametrization with $\gamma_i=0$ in Eq.~(\ref{eq:ff-input}) would
introduce artificial correlations between the behavior of fragmentation functions
in different regions of $z$ obscuring also the assessment of uncertainties.
We find that the extra term $\sim(1-z)^{\delta_i}$ in Eq.~(\ref{eq:ff-input})
considerably improves the quality of the global fit, closely related to the fact that
the analysis of fragmentation functions is restricted to medium-to-large $z$.
Accordingly, additional power terms in $z$, emphasizing the small $z$ region,
have little or no impact on the fit and are not pursued further.
The initial scale $\mu_0$ for the $Q^2$-evolution is taken to be $\mu_0=1\,\mathrm{GeV}$
in our analysis.

Since the initial fragmentation functions (\ref{eq:ff-input}) at scale $\mu_0$ should
not involve more free parameters than can be extracted from data, we have to
impose, however, certain relations upon the individual fragmentation functions
for pions and kaons.
We have checked in each case that relaxing these assumptions indeed does not
significantly improve the $\chi^2$ of the fit of presently available data
to warrant any additional parameters.
In detail, for $\{u,\,\bar{u},\,d,\,\bar{d}\}\to \pi^+$ we impose
isospin symmetry for the sea fragmentation functions, i.e.,
\begin{equation}
\label{eq:iso}
D_{\bar{u}}^{\pi^+}=D_{d}^{\pi^+},
\end{equation}
but we allow for slightly different normalizations in the $q+\bar{q}$ sum:
\begin{equation}
\label{eq:val_break}
D_{d+\bar{d}}^{\pi^+}= N D_{u+\bar{u}}^{\pi^+}.
\end{equation}
For strange quarks it is assumed that
\begin{equation}
\label{eq:sea_break}
D_s^{\pi^+}=D_{\bar{s}}^{\pi^+} =N^{\prime} D_{\bar{u}}^{\pi^+}
\end{equation}
with
$N^{\prime}$ independent of $z$.

It is worth noticing that {\em assuming} $N=N^{\prime}=1$ \cite{ref:kretzer,ref:hirai}
in Eqs.~(\ref{eq:val_break}) and (\ref{eq:sea_break}), respectively, SIA data alone
allow to distinguish  between favored and unfavored fragmentation functions
in principle.
We shall scrutinize the compatibility of these assumptions with SIDIS and
hadronic scattering data in Sec.~\ref{sec:unc-results}. At any rate, their impact
on the assessment of uncertainties of fragmentation functions is highly non trivial.

For charged kaons we fit $D_{u+\bar{u}}^{K^+}$ and
$D_{s+\bar{s}}^{K^+}$ independently to account for the phenomenological
expectation that the formation of secondary $s\bar{s}$ pairs,
which is required to form a $|K^+\rangle=|u\bar{s}\rangle$ from a $u$ but not from an $\bar{s}$ quark,
should be suppressed.
Indeed, we find from our fit, see Sec. IV below, that
$D_{s+\bar{s}}^{K^+}> D_{u+\bar{u}}^{K^+}$ in line with that expectation.
For the unfavored fragmentation the data are unable to discriminate between
flavors and, consequently, we assume that all distributions have the
same functional form:
\begin{equation}
\label{eq:sea_ka}
D_{\bar{u}}^{K^+}=D_{s}^{K^+}=D_d^{K^+}=D_{\bar{d}}^{K^+}.
\end{equation}

We adopt the functional form (\ref{eq:ff-input}) also for the
fragmentation of heavy charm and bottom quarks into charged pions
and kaons but setting $\gamma_i=0$. As in \cite{ref:kretzer,ref:kkp,ref:akk,ref:hirai}
we assume that $D_{c}^{H}=D_{\bar{c}}^{H}$ and $D_{b}^{H}=D_{\bar{b}}^{H}$
for $H=\pi^+,\, K^+$.
Heavy flavors are included discontinuously as massless partons
in the evolution (\ref{eq:singevol})
above their $\overline{\text{MS}}$ ``thresholds'', $Q=m_{c,b}$,
with $m_{c,b}$ denoting the mass of the charm and bottom quark,
respectively.
This treatment of heavy flavors is very much at variance
with heavy quark parton densities, where very elaborate schemes have been
developed to properly include mass effects near threshold and to
resum large logarithms $\sim \ln m_{c,b}^2/Q^2$ for $Q^2\gg m_{c,b}^2$.
Only SIA data at $\sqrt{s}\gg m_{c,b}$ are sensitive, however, to charm and bottom
fragmentation in the analysis. Neither the charged pion or kaon
multiplicities in SIDIS nor hadron production data from proton-proton
collisions at RHIC receive any noticeable contribution from heavy quark fragmentation.
Therefore the massless approximation outlined above,
also adopted in \cite{ref:kretzer,ref:kkp,ref:akk,ref:hirai}, appears
to be sufficient for the time being. However, we note that a dynamical,
parameter-free generation of the heavy flavor component to light
meson fragmentation functions based on NLO matching conditions has been developed
recently in \cite{ref:cacciari}. This might prove to be a viable alternative
to the presently adopted framework in the future.

Thus in total we have to determine 23 (24) parameters in the global $\chi^2$
analysis describing the hadronization of quarks and gluons
into positively charged pions (kaons).
Corresponding fragmentation functions into $\pi^-$, $K^-$ are
obtained as usual by charge conjugation and those for neutral pions
by assuming $D_i^{\pi^0}= [D_i^{\pi^+}+D_i^{\pi^-}]/2$.

We numerically solve the renormalization group equation in NLO
\begin{equation}
\label{eq:alphas}
\frac{d\alpha_s(\mu^2)}{d\ln \mu^2}=
-\frac{\beta_0}{4\pi} \alpha_s^2(\mu^2)-
\frac{\beta_1}{(4\pi)^2} \alpha_s^3(\mu^2)
\end{equation}
with $\beta_0=11-2 n_f/3$ and $\beta_1=102-38 n_f/3$
to determine the running of the strong coupling.
The number of active flavors $n_f$ is increased upon crossing the
heavy flavor thresholds at $\mu^2=m_c^2$ and $\mu^2=m_b^2$
for which we choose \cite{ref:mrst} $m_c=1.43\,\text{GeV} $ and $m_b=4.3\,\text{GeV}$.
We specify $\alpha_s$ in the solution of (\ref{eq:alphas}) for $n_f=4$ by adopting
\cite{ref:mrst} $\Lambda_{\text{QCD}}^{(n_f=4)}=334\,\text{MeV}$.
For our leading order (LO) analysis we use \cite{ref:mrstlo}
$\Lambda_{\text{QCD}}^{(n_f=4)}=220\,\text{MeV}$ and, of course, set $\beta_1=0$ in
Eq.~(\ref{eq:alphas}).
%
\subsection{\label{sec:datasets} Selection of Data Sets}
%
The parameters describing the fragmentation functions for
pions and kaons at scale $\mu_0$ in Eq.~(\ref{eq:ff-input})
are determined by a standard $\chi^2$ minimization
for $N$ data points, where
\begin{equation}
\label{eq:chi2}
\chi^2=\sum_{i=1}^N \frac{(T_i-E_i)^2}{\delta E_i^2},
\end{equation}
$E_i$ is the measured value of a given observable,
$\delta E_i$ the error associated with this measurement, and
$T_i$ is the corresponding theoretical estimate for a
given set of parameters in (\ref{eq:ff-input}).
Since the full error correlation matrices are not available for
most of the data entering the global analysis, we
take, as usual \cite{ref:kretzer,ref:kkp,ref:akk,ref:hirai},
the statistical and systematical errors in quadrature
in $\delta E_i$.

In (\ref{eq:chi2}) we use charged pion and kaon production
data in SIA from TPC \cite{ref:tpcdata} at $\sqrt{s}=29\,\mathrm{GeV}$,
SLD \cite{ref:slddata}, ALEPH \cite{ref:alephdata}, DELPHI \cite{ref:delphidata},
and  OPAL \cite{ref:opaldata}, all at $\sqrt{s}=M_Z$.
To further constrain fragmentation functions through scale evolution,
we also use data from TASSO \cite{ref:tassodata} at intermediate
c.m.s.\ energies of $\sqrt{s}=33$ and
$44\,\mathrm{GeV}$, which suffer, however, from rather large experimental
uncertainties as compared to the other SIA data listed above.
Other measurements of SIA \cite{ref:whalley} have too large experimental uncertainties and
hence are not used in our analysis.
Because of the conceptual problem with fragmentation functions
at small $z$ outlined in Sec.~\ref{sec:ffs},
the cut $z_{\min}=0.05$ (0.1) is imposed for all
pion (kaon) data sets.

Besides these fully inclusive measurements also ``flavor tagged''
SIA results are available, where the quark flavor refers to the
primary $q\bar{q}$ pair created by the intermediate photon or $Z$-boson.
ALEPH \cite{ref:alephdata}, DELPHI \cite{ref:delphidata}, and TPC \cite{ref:tpcdata}
provide tagged results distinguishing between the sum of
light $u,\, d,\, s$ quarks, charm, and bottom events. This information
is of particular value for the flavor decomposition,
as the fully inclusive, ``untagged'' data mainly
constrain the flavor singlet combination $D_{\Sigma}^H(z)$ on the $Z$-resonance
due overwhelming statistical precision of the LEP and SLD data
and the fact that $\hat{e}_u^2\simeq \hat{e}_d^2$ at $Q=M_Z$.
On the downside, flavor tagged results can neither be measured directly nor can
they be unambiguously interpreted and calculated in pQCD. Flavor enriched
samples are unfolded based on Monte-Carlo simulations estimating the
flavor composition of the data sets. For heavy flavor tagged data,
a further complication arises due to possible contaminations
by weak decay channels.
In our analysis, we obtain the corresponding theoretical results $T_i$
by summing in Eqs.~(\ref{eq:ee-xsec})-(\ref{eq:flnlo}) only over those
flavors which are tagged experimentally. At NLO this accounts for
gluon radiation as well as the possibility that not the original
(anti-)quark but the radiated gluon produces the observed hadron.

In addition to the flavor tagged results just discussed,
OPAL \cite{ref:opaleta} has presented fully flavor separated ``data''
in terms of ``probabilities'' $\eta_i^H(x_p,s)$ for a quark flavor
$i=q+\bar{q}$ to produce a ``jet'' containing the hadron $H$ with
a momentum fraction $z$ larger than $x_p$.
Needless to say, that these results are even more difficult to interpret within
pQCD beyond the LO and should not be taken to literally.
To take this into account we assign an up to $10\%$ normalization uncertainty
to the OPAL tagging probabilities $\eta_i^H$  \cite{ref:opaleta} in the fit.
Nevertheless some tension with other data sets remains, in particular for
$\eta_c^H$ and $\eta_b^H$, as will be discussed below.
In our analysis we interpret the OPAL results as
\begin{equation}
\label{eq:opaleta}
\eta_i^H(x_p,Q=M_Z) = \int_{x_p}^1 dz \frac{1}{\sigma_{tot}}
\frac{d\sigma^H}{dz}\Bigg|_{i=q},
\end{equation}
where the subscript $i=q$ denotes that in Eqs.~(\ref{eq:ee-xsec})-(\ref{eq:flnlo})
all sums only include the specific quark flavor $i$.

To further constrain the fragmentation of different flavors, as well as too
separate favored (valence) and unfavored (sea) fragmentation,
we include experimental information from SIDIS, see Sec.~\ref{sec:sidis}.
More specifically, we make use of (preliminary) charged pion and kaon multiplicities
from the HERMES experiment \cite{ref:hermessidis}.
These data also provide an important consistency check of the flavor
separation obtained from flavor tagged SIA ``data'', as well as of pQCD
scale evolution (\ref{eq:singevol}) since they refer to much lower scales
$\mu\simeq Q = 1 \div 3\,\mathrm{GeV}\ll M_Z$.

In the $\chi^2$ minimization we have to account for the fact that
the SIDIS data are taken in certain bins of $z$ and $Q^2$ \cite{ref:hermessidis},
whereas the theoretical estimates (\ref{eq:sidis-xsec}) are computed for the
center of each bin. We estimate the corresponding uncertainty as the
maximal variation of the cross section within each particular bin,
see Figs.~\ref{fig:sidis-pion} and \ref{fig:sidis-kaon} below,
and add it in quadrature to $\delta E_i$ in Eq.~(\ref{eq:chi2}).
In case of charged kaon multiplicities, we allow for an additional
5\% uncertainty to account for a possible inadequacy of the massless
approximation.

A wealth of new data on single-inclusive hadron
production from RHIC experiments
\cite{ref:phenixpion,ref:starpion,ref:brahms,ref:starkaon} 
have also been  included in our global analysis.
These encompass the $p_T$ spectrum of neutral pions at central
rapidities $|\eta|\le 0.35$ by PHENIX \cite{ref:phenixpion} and
at three different forward rapidities $\langle \eta\rangle=3.3$, 3.8, and 4.0
by STAR \cite{ref:starpion}. For the latter we
exclude the most forward bin, $\langle \eta \rangle =4.0$, from the
fit as it has large theoretical uncertainties due the small $p_T$
values probed. BRAHMS has very recently
published $p_T$ spectra for identified charged pions and kaons
at two values of (forward) rapidities $\eta=2.95$ and $3.3$ \cite{ref:brahms}
of which we use only the former in the fit for similar reasons as above
for STAR. In addition, there are data on $K_S^0$ production at central
rapidities $|\eta|\le 0.5$ from STAR \cite{ref:starkaon}.
To accommodate the $K_S^0$ data in the fit, we assume that
$K^0_S=(K^+ + K^-)/2$ with  $u\to K^+$ and $d\to K^+$ fragmentation functions
interchanged.

For all hadronic data from RHIC, an additional 5\% error is assigned
in quadrature to $\delta E_i$ in evaluations of $\chi^2$ in Eq.~(\ref{eq:chi2})
as a rather conservative estimate of the theoretical uncertainties related to
the choice of the factorization and renormalization scales in (\ref{eq:pp-xsec}).

To ease possible tensions between certain data sets, we allow the data
to ``float'' within the normalization uncertainties quoted by each experiment.
More precisely, in addition to the ${\cal{O}}(20)$ parameters describing the fragmentation functions
in (\ref{eq:ff-input}), we also fit a set of relative normalization factors for each experiment
in the $\chi^2$ minimization to determine the optimum fit.
We note that the possibility of normalization uncertainties has been not
addressed in all previous analysis of SIA data \cite{ref:kretzer,ref:kkp,ref:akk,ref:hirai}.

\subsection{\label{sec:mellin}Mellin Technique}
%
The integro-differential evolution equations (\ref{eq:singevol})
can be straightforwardly solved {\em analytically} in Mellin $n$-moment
space along the lines described, e.g., in Ref.~\cite{ref:grv1}.
The Mellin moments of, for instance, the fragmentation functions
$D_i^H(z,Q^2)$, are defined as
\begin{equation}
\label{eq:mellin-moment}
D_i^{H}(n,Q^2) \equiv \int_0^1 dz \;z^{n-1} D_i^H(z,Q^2),
\end{equation}
and can be expressed in terms of Euler Beta functions for our
ansatz (\ref{eq:ff-input}) at scale $\mu_0$.
The relevant moments of the evolution kernels
$P_{ij}^{(T)}(n)$ are given in \cite{ref:grv2}.
The evolved fragmentation functions in $z$-space are
re-obtained by an inverse Mellin transform given by
\begin{equation}
\label{eq:mellin-inverse}
D_i^H(z,Q^2) = \frac{1}{2 \pi i} \int_{{\cal C}_n} dn \;
z^{-n} D_i^H(n,Q^2),
\end{equation}
where ${\cal C}_n$ denotes an appropriately chosen
contour in the complex $n$ plane.

The property that numerically very time-consuming convolutions
in $z$-space factorize into simple products under Mellin moments
makes them also an ideal tool to compute cross sections.
Also ``plus distributions'', which regularize singularities
as $z\to 1$, are much easier to handle.
For the SIA cross section (\ref{eq:ee-xsec}) - (\ref{eq:flnlo})
the virtue of Mellin moments is immediately obvious since they
can be taken analytically for the hard scattering coefficient
functions $C_{q,g}^{1,L}$  and can be found in Ref.~\cite{ref:grv2}.
We note that also the ``tagging probabilities'' $\eta_i^H$ in
Eq.~(\ref{eq:opaleta}) as obtained by OPAL \cite{ref:opaleta}
can be straightforwardly computed in Mellin moment space
\begin{equation}
\label{eq:opal-mellin}
\eta_i^H(x_p,Q=M_Z) = \frac{1}{2\pi i} \int_{{\cal{C}}_n} dn
\frac{1-x_p^{(1-n)}}{1-n} \frac{1}{\sigma_{tot}}
\frac{d\sigma^H}{dn} \Bigg|_{i=q},
\end{equation}
where $\frac{1}{\sigma_{tot}}\frac{d\sigma^H}{dn}\left|_{i=q} \right.$ denote
the Mellin moments of the SIA cross section (\ref{eq:ee-xsec}) - (\ref{eq:flnlo})
for a single flavor $i=q$.

The direct use of the SIDIS cross section (\ref{eq:sidis-xsec})-(\ref{eq:flsidis})
in our global analysis would be rather time-consuming and awkward
though not impossible since the partonic coefficient functions $C^{1,L}_{ij}$
 are still fairly simple.
Again, transforming Eqs.~(\ref{eq:sidis-xsec})-(\ref{eq:flsidis}) to
Mellin space is much more appropriate in extensive numerical analyses.
As for SIA, the Mellin moments of the coefficient functions
can be taken completely analytically and can be found in \cite{ref:aempi,ref:mellin}.
Since the $C^{1,L}_{ij}$ depend both on $x$ and $z$, a double Mellin
transform is required for SIDIS:
\begin{equation}
\label{eq:sidis-mellin}
C_{ij}^{(1,L)}(n,m) \equiv
\int_0^1 dx\, x^{n-1} \int_0^1 dz\, z^{m-1}
C_{ij}^{(1,L)}(x,z),
\end{equation}
where the dependence on the factorization and renormalization scales is suppressed
in (\ref{eq:sidis-mellin}). Upon combining the $C_{ij}^{(1,L)}(n,m)$
with the appropriate $n$ and $m$ moments of the evolved parton densities
$f_i(n,Q^2)$ and fragmentation functions $D_j^H(m,Q^2)$,
respectively, the SIDIS cross section is obtained by a numerical fast
double inverse Mellin transform as described in \cite{ref:mellin}.

For the much more complex and lengthy partonic hard cross sections
entering a calculation of hadronic cross sections, Eq.~(\ref{eq:pp-xsec}),
at NLO accuracy,
Mellin moments can no longer be computed analytically. Nevertheless
it is in the analysis of hadron-hadron collision data where the Mellin
moment technique exhibits its full potential and usefulness
\cite{ref:mellin,ref:mellin2}.
The crucial, but simple, ``trick'' in applying Mellin moments to
Eq.~(\ref{eq:pp-xsec}) is to express the fragmentation functions $D_c^H(z)$
by their Mellin inverses $D_c^H(n)$ defined in Eq.~(\ref{eq:mellin-inverse}).
One subsequently interchanges integrations and arrives schematically at
\begin{eqnarray}
\label{eq:pp-mellin}
\nonumber
E_H \frac{d^3 \sigma}{dp_H^3} &=& \frac{1}{2\pi i} \sum_{c}
\int_{{\cal{C}}_n} dn D_c^H(n) \\
&&\times \left[\sum_{ab}
f_a \otimes f_b  \otimes d\hat{\sigma}_{ab}^{c} \otimes z^{-n} \right].
\end{eqnarray}
One can now pre-calculate the quantities $
d\tilde{\hat{\sigma}}_{ab}^c(n)\equiv
\sum_{ab} f_a \otimes f_b  \otimes d\hat{\sigma}_{ab}^{c}
\otimes z^{-n}$
in (\ref{eq:pp-mellin}), which do not depend on the
fragmentation functions $D_c^H(n)$,
{\em prior} to the fit for each contributing combination
of single-inclusive subprocesses producing a certain parton
$c$ and in each experimental bin.
We emphasize that in this way all the tedious and time-consuming integrations
are already dealt with.

The inverse Mellin transformation, which finally links the
moments of the fitted fragmentation functions with the pre-calculated
$d\tilde{\hat{\sigma}}_{ab}^c(n)$, of course still needs to be performed
in each step of the fitting procedure. However,
the integration over $n$ is extremely fast to perform by
choosing the values for $n$ on the contour ${\cal C}_n$ in (\ref{eq:pp-mellin})
simply as the supports for a Gaussian integration.
The point here is that the integrand in $n$ falls off very
rapidly as $|n|$ increases along the contour for two reasons:
first, each fragmentation function is expected to fall off at
least as a power $(1-z)^{\beta}$, $\beta\geq 1$, at large $z$,
which in moment space converts into a fall-off of
$\sim 1/n^{\beta+1}$ or higher.
Secondly, we may choose a contour ${\cal{C}}_n$ that is bent by an
angle with respect to the standard vertical direction such that for large
$|n|$, $(z)^{-n}$ decreases exponentially along the contour \cite{ref:mellin}.
This greatly improves the numerical convergence of the calculation
of the $d\tilde{\hat{\sigma}}_{ab}^c(n)$ in Eq.~(\ref{eq:pp-mellin})
and also gives them a rapid fall-off at large arguments.
We note that for all practical purposes of our global analysis
between 64 and 92 $n$ moments of $d\tilde{\hat{\sigma}}_{ab}^c(n)$ are sufficient
to reproduce the cross section (\ref{eq:pp-xsec})
to an accuracy of much better than $1\%$ for all data points used in the fit.

The crucial asset of the Mellin method is the speed at which one
can calculate the full hadronic cross section at NLO
{\em without} approximations, once the moments
$d\tilde{\hat{\sigma}}_{ab}^c(n)$ have been pre-calculated.
To give an example, a full NLO computation of all 78
data points from RHIC used in the analysis of the pion
fragmentation functions takes much less than 0.1 second as compared
to several minutes using Eq.~(\ref{eq:pp-xsec}) directly.
Since a few thousand evaluations of each data point are required in course of
the $\chi^2$ minimization, this clearly shows the value of
using Mellin moments.

\subsection{\label{sec:lagrange}Uncertainties: Lagrange Multiplier Technique}
%
The most difficult but crucial issue to be addressed in a
global analysis is the estimate of the uncertainties in the
extraction of the individual fragmentation functions $D_i^H$.
Without a proper assessment of errors, any interpretation of
the results of the fit or predictions for observables
based on the fitted $D_i^H$ are incomplete and perhaps even misleading.
Uncertainties in global analyses have been thoroughly studied in the context of unpolarized
parton distributions (PDFs) \cite{ref:BOTJE,ref:Stump:2001gu,ref:mrst},
where the number and precision of the data available
is much more significant. A reliable estimate of the
errors arising from PDFs in predictions for observables
related to, e.g., new physics or Higgs boson searches at CERN-LHC
is of utmost importance.

The possible sources of uncertainties for parton densities or
fragmentation functions can be classified into those associated
with experimental errors on the data, and those related to
theoretical or phenomenological assumptions in the global fitting procedure.
The latter include, for example, higher order QCD effects in the analyzed
cross sections and their scale dependence, the particular choice of the
parametric form of the distributions at the initial scale, and other model
assumptions such as flavor and charge conjugation symmetries.
Clearly, while the first category is usually under control,
the second one is particularly difficult to quantify.

Many strategies have been conceived and explored in order to assess
the uncertainties of PDFs and their propagation to observables,
specially those associated with experimental errors in the data.
These include the ``Hessian approach'' \cite{ref:Stump:2001gu},
which {\em assumes} that the deviation in $\chi^2$ for the global fit is
quadratic in the parameters specifying the input PDFs away from
their optimum fit values. Then one propagates these uncertainties
of PDFs {\em linearly} to observables. Alternatively, the
``Lagrange multiplier method'' \cite{ref:Stump:2001gu} probes
the uncertainty in any observable or quantity of interest much more directly.
It relates the range of variation of one or more physical observables
dependent upon PDFs to the variation in the $\chi^2$ used to
judge the goodness of the fit to data.
Specifically, it can be implemented by minimizing the function
\begin{equation}
\label{eq:lagrange}
\Phi(\lambda_i,\{a_j\})=\chi^2(\{a_j\})+\sum_i \lambda_i\,O_i(\{a_j\})
\end{equation}
with respect to the set of parameters $\{a_j\}$ describing the PDFs,
for fixed values of the Lagrange multipliers $\lambda_i$.
Each one of the parameters $\lambda_i$ is related to
an observable $O_i$ depending on $\{a_j\}$.
The choice $\lambda_i=0$ in (\ref{eq:lagrange})
corresponds to the optimum global fit $\{a_j\}$,
for which $\chi^2(\{a_j\})\equiv\chi^2_0$ and $O_i(\{a_j\})\equiv O_i^0$.
Minimizing $\Phi(\lambda_i, \{a_j\})$ for $\lambda_i\neq 0$
deteriorates the quality of the fit to data and other values for the
observable  $O_i(\{a_j\})$ are found from the set
of newly fitted parameters $\{a_j\}$.
From a series of global fits for different values of $\lambda_i$,
the $\chi^2(\{a_j\})$ profile depending on different values of $O_i$
can be mapped out.
In other words, this tell us by how much the fit to
data deteriorates if we force the PDFs to yield a prediction for an
observable different to the one obtained with the best fit $O_i^0$.

The value and practical feasibility of the Lagrange multiplier technique has been
demonstrated not only for the highly sophisticated global analyses
of unpolarized PDFs \cite{ref:Stump:2001gu}, but also in case of
polarized PDFs \cite{ref:polpdf}.
In Sec.~\ref{sec:unc-results}, we will show that the
same holds for the analysis of fragmentation functions.
Here, the limitations due to the available data are in some sense similar to
those we encounter for polarized PDFs: the bulk of the data
(SIA and spin-dependent DIS in case of fragmentation functions and
polarized PDFs, respectively)
neither determine the gluon well nor allow for a reliable flavor separation.
In both types of analysis, SIDIS and hadronic data provide invaluable
constraints on the parameter space describing the input densities.

In an {\em ideal} situation, where every source of uncertainty is
well understood and fully accounted for, all $\chi^2$ profiles,
including those for the parameters $\{a_j\}$ of the fit,
would be parabolic, and the 1-$\sigma$ uncertainty for any observable would
correspond to a change in $\chi^2$ by one unit, i.e., $\Delta \chi^2=1$.
This is, of course, rarely the case, and in order to account for
missing correlated experimental errors or theoretical uncertainties
in global analysis it is customary to consider instead of
$\Delta \chi^2=1$ a $2\div 5\%$ variation in $\chi^2$ as a more
conservative estimate of the range of uncertainty
\cite{ref:BOTJE,ref:Stump:2001gu,ref:mrst,ref:polpdf}.

In addition to the possibility of assessing the uncertainties of
parameters $a_j$ or observables $O_i(\{a_j\})$, the Lagrange multiplier
method allows to elucidate the role of each subset of data included in the fit
in constraining a certain quantity. One just needs to determine
the shape and variation of the partial contribution $\Delta \chi^2_n$
of a particular subset $n$ of data to the total $\chi^2$ as the observable
changes depending on the Lagrange multipliers.
When a given subset of data can by itself determine, say, a given observable
$O_i(\{a_j\})$, the profile of $\Delta \chi^2_n$ w.r.t.\  $O_i(\{a_j\})$
is expected to be roughly parabolic, with a minimum close to the
``preferred'' value $O_i^0$ determined by the optimum global fit.
However, when a given subset of data does not fully constrain the observable
$O_i(\{a_j\})$, its profile w.r.t.\  $O_i(\{a_j\})$
is either flat or increases (decreases) monotonically without minimum
in the range of variation of the observable.
In general, constraints on $O_i(\{a_j\})$ in a global analysis
result from the subtle interplay of several subsets of data.
The combination of the different partial contributions to $\chi^2$,
even of those that by themselves do not show a minimum,
define the final $\chi^2$ profile and the best fit value $O_i^0$,
thereby highlighting the complementary nature of a global analysis.
For our fragmentation functions this will be illustrated in detail in
Sec.~\ref{sec:unc-results}.

\section{\label{sec:results} Results}
%
In this Section we discuss in detail the results of our global
analysis of fragmentation functions for pions and kaons.
We present the parameters of the optimum fits
describing the $D_i^{\pi^+,K^+}$ at the input scale $\mu_0$,
compare to the data used in the analysis, and give
$\chi^2$ values for each individual set of data used.
Detailed comparisons are made with the results obtained in
the analyses of SIA data in Refs. \cite{ref:kretzer} and
\cite{ref:akk}, in the following labeled as KRE and AKK,
respectively.
Even though we are mainly interested in a precise extraction
of fragmentation functions at NLO accuracy,
we also briefly present corresponding results
of a global analysis performed at LO approximation.
The significantly better $\chi^2$ of the NLO sets
highlights the importance of the NLO corrections and the
limitations of a LO analysis.
Nevertheless, our LO sets should be used in calculations of observables
where NLO corrections are not available, or in event generators
limited to LO accuracy.

\subsection{\label{sec:nlopion}NLO analysis of pion fragmentation functions}
\begin{figure*}[h!]
\begin{center}
\vspace*{-0.6cm}
\epsfig{figure=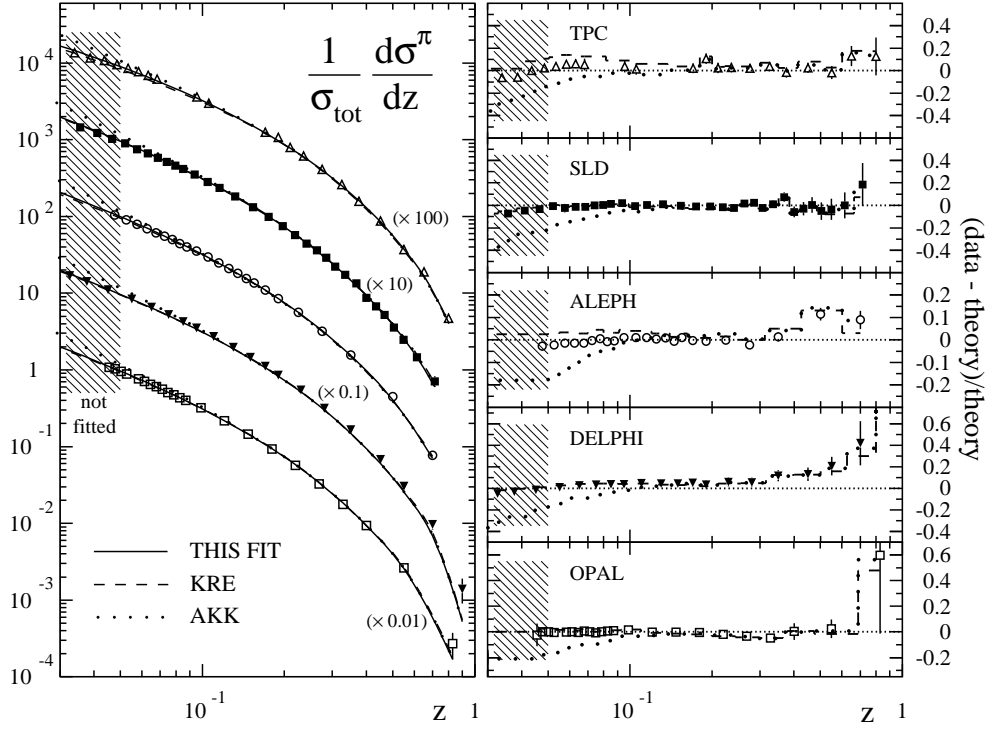,width=0.75\textwidth}
\end{center}
\vspace*{-0.7cm}
\caption{{\bf L.h.s.} comparison of our NLO results for $(1/\sigma_{tot}) d\sigma^{\pi}/dz$ according to
Eq.~(\ref{eq:ee-xsec}) with the data sets for inclusive pion production in SIA used in the fit, see
Tab.\ \ref{tab:exppiontab}. {\bf R.h.s.} ``(data-theory)/theory'' for our NLO results
for each of the data sets. Also shown are the results obtained with the KRE \cite{ref:kretzer}
and AKK \cite{ref:akk} parameterizations, dashed and dotted lines, respectively.
\label{fig:pioninclusive}}
\vspace*{-0.5cm}
\end{figure*}

\begin{figure*}[b!]
\begin{center}
\vspace*{-0.6cm}
\epsfig{figure=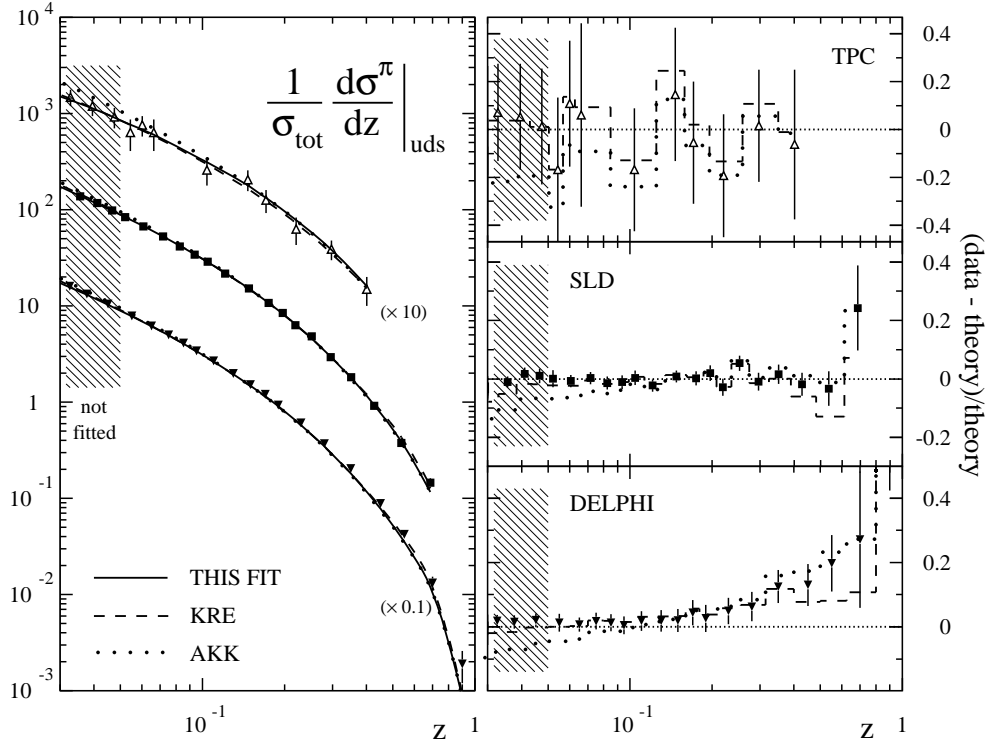,width=0.75\textwidth}
\end{center}
\vspace*{-0.7cm}
\caption{Same as in Fig.\ \ref{fig:pioninclusive} but now for light quark (``uds'') tagged
cross sections.
\label{fig:pion-uds}}
\vspace*{-0.5cm}
\end{figure*}

\begin{figure*}[h!]
\begin{center}
\vspace*{-0.6cm}
\epsfig{figure=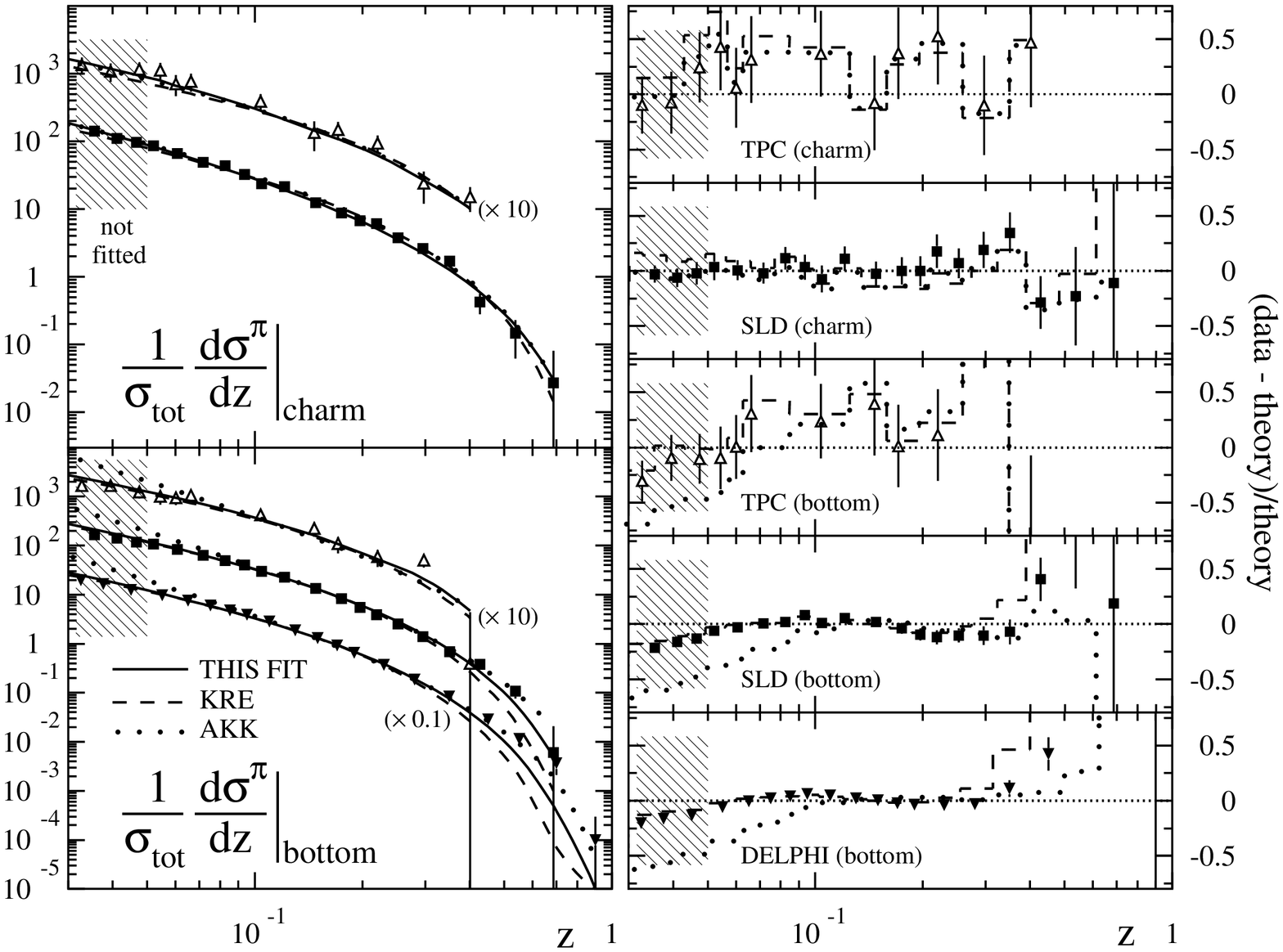,width=0.75\textwidth}
\end{center}
\vspace*{-0.7cm}
\caption{Same as in Fig.\ \ref{fig:pioninclusive} but now for charm
and bottom quark tagged cross sections.
\label{fig:pion-cb}}
\vspace*{-0.5cm}
\end{figure*}

\begin{figure*}[b!]
\begin{center}
\vspace*{-0.2cm}
\epsfig{figure=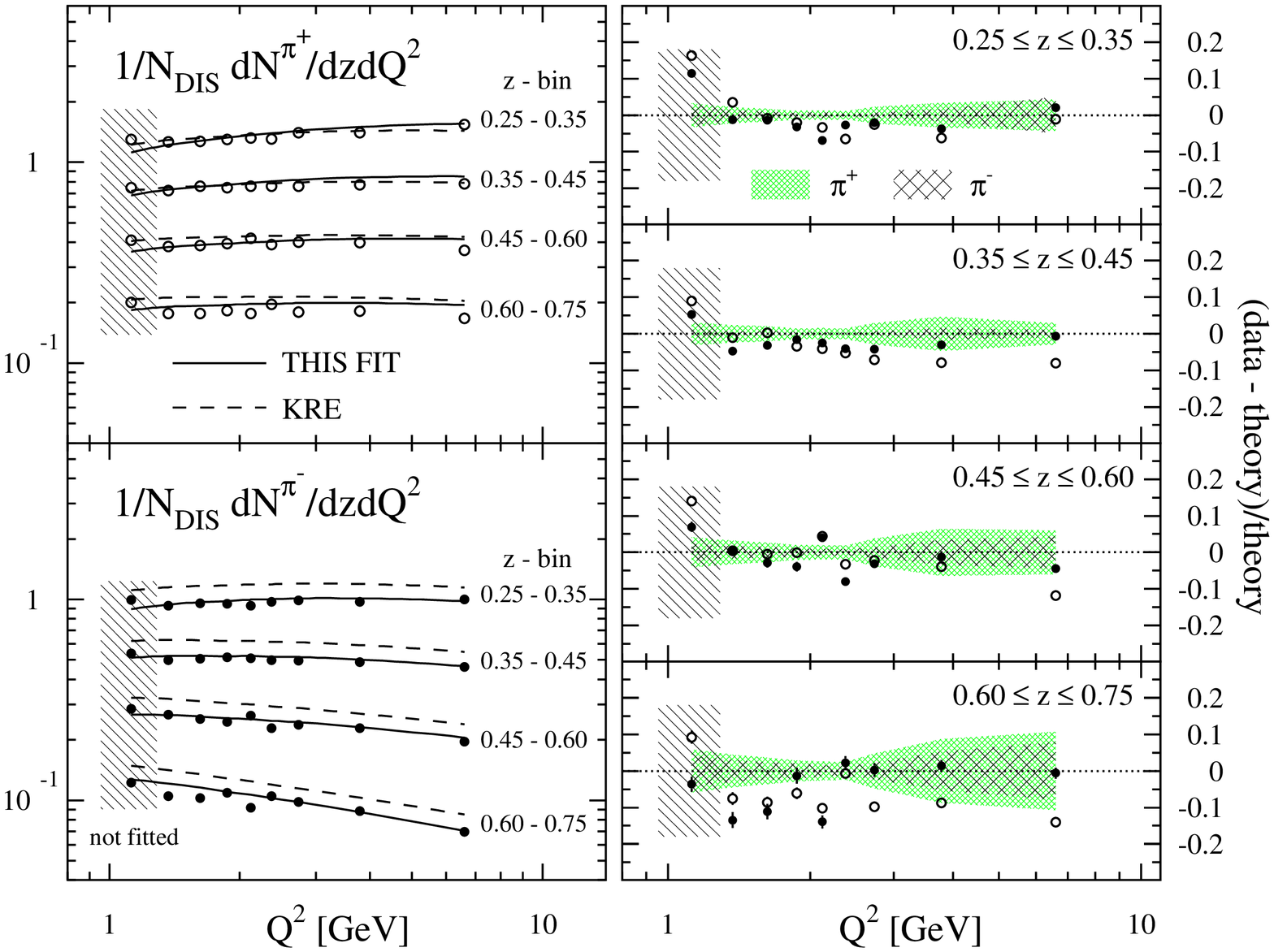,width=0.75\textwidth}
\end{center}
\vspace*{-0.7cm}
\caption{{\bf L.h.s.} comparison of our NLO results for
charged pion multiplicities in SIDIS, $(1/N_{DIS}) dN^{\pi^{\pm}}/dzdQ^2$,
with preliminary HERMES data \cite{ref:hermessidis}.
Also shown are the results obtained with the KRE \cite{ref:kretzer} parametrization.
{\bf R.h.s.} ``(data-theory)/theory'' for our NLO results, open and full
circles denote $\pi^+$ and $\pi^-$ multiplicities, respectively.
The shaded bands indicate estimates of theoretical uncertainties due to finite
bin-size effects (see text).
\label{fig:sidis-pion}}
\vspace*{-0.5cm}
\end{figure*}

\begin{figure*}[t]
\begin{center}
\vspace*{-0.6cm}
\epsfig{figure=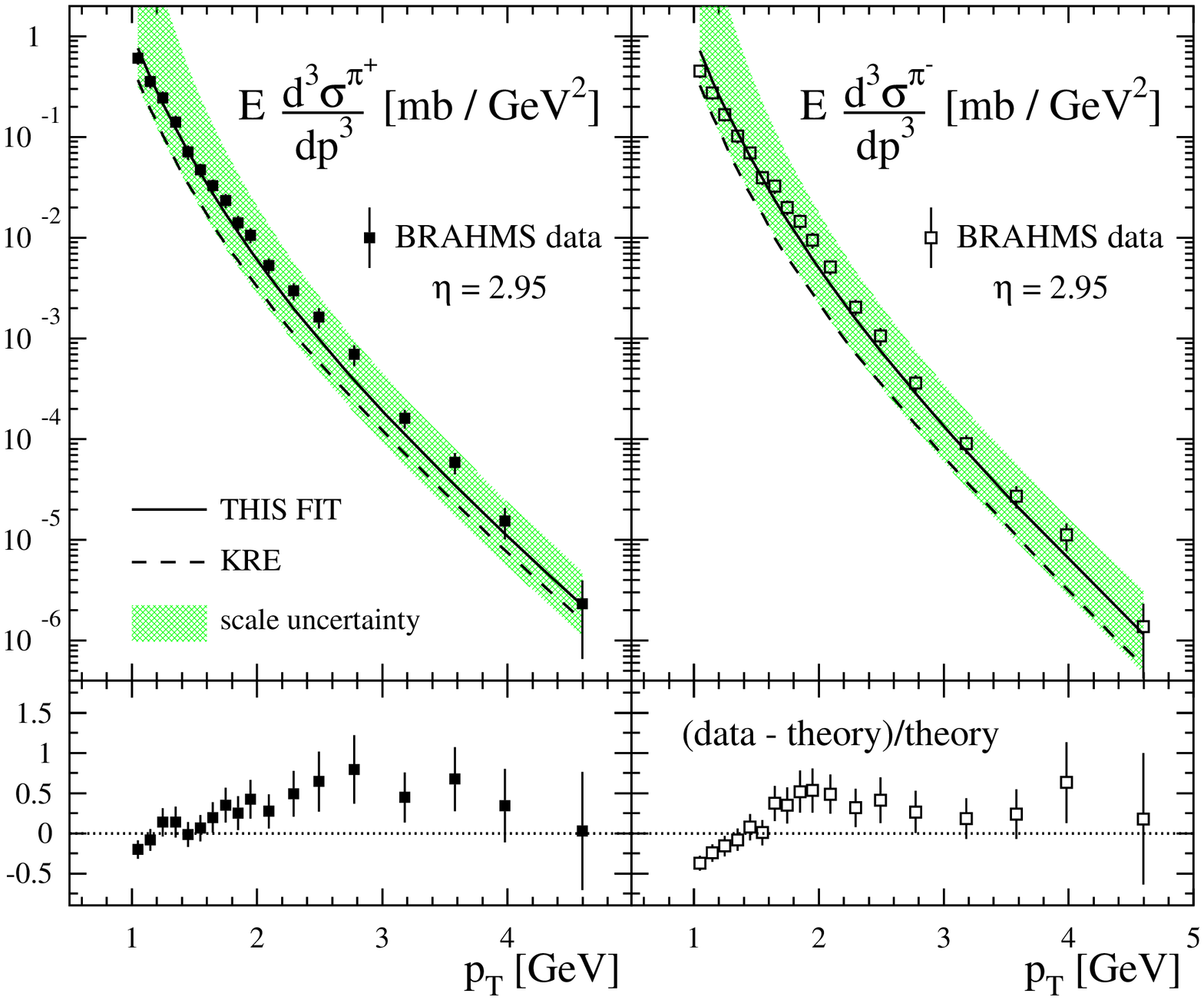,width=0.75\textwidth}
\end{center}
\vspace*{-0.7cm}
\caption{{\bf upper panels} comparison of our NLO results for
single-inclusive charged pion production $pp\rightarrow \pi^{\pm} X$
at rapidity $\eta=2.95$  (solid lines) with BRAHMS data \cite{ref:brahms} using $\mu_f=\mu_r=p_T$.
Also shown are the results obtained with the KRE \cite{ref:kretzer} parametrization (dashed lines).
The shaded bands indicate theoretical uncertainties when all
scales are varied in the range $p_T/2\le \mu_f=\mu_r \le 2 p_T$.
{\bf lower panels} ``(data-theory)/theory'' for our NLO results.
\label{fig:brahms-pion}}
\vspace*{-0.5cm}
\end{figure*}

\begin{figure}[thb!]
\begin{center}
\vspace*{-0.6cm}
\epsfig{figure=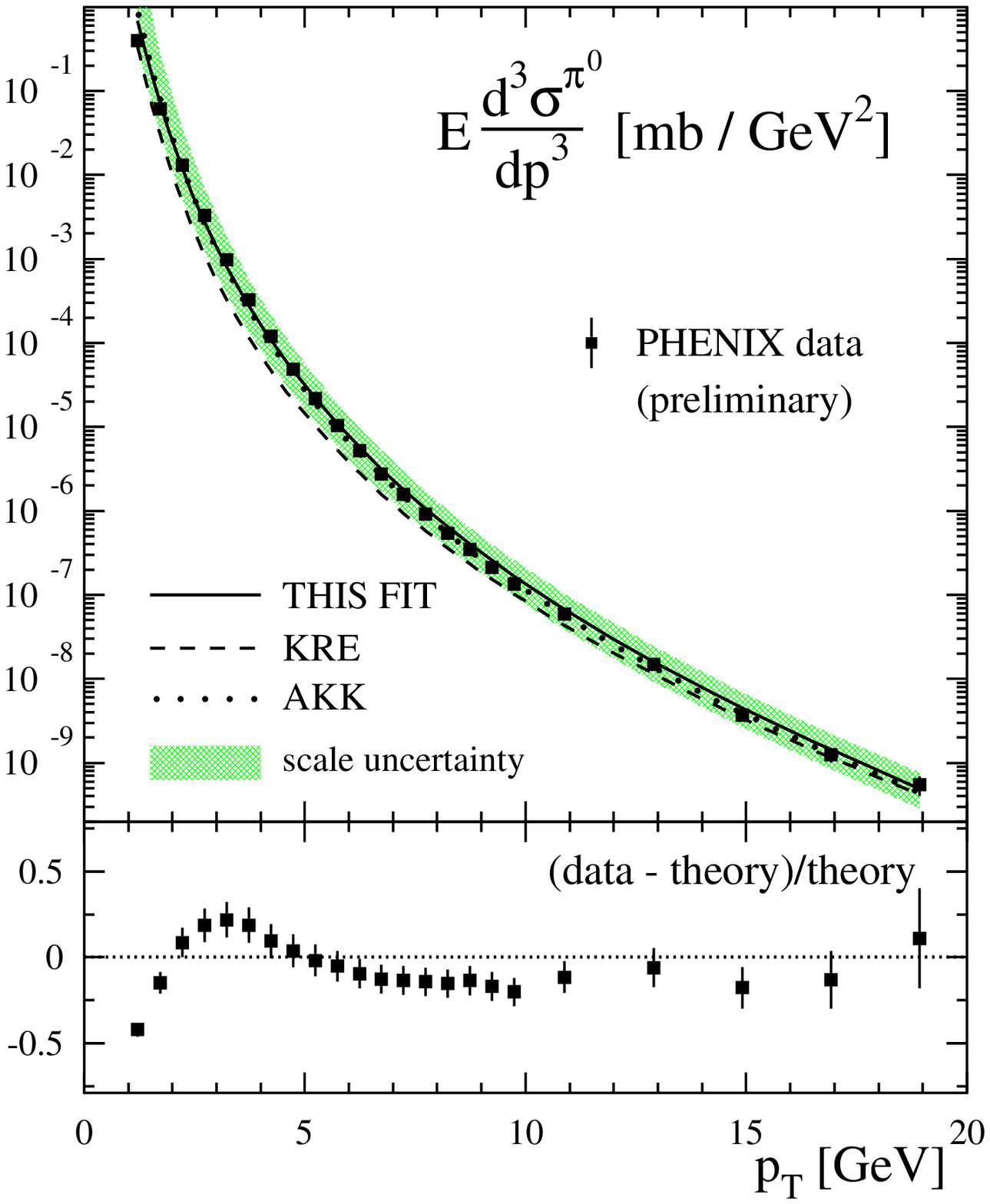,width=0.495\textwidth}
\end{center}
\vspace*{-0.7cm}
\caption{Same as in Fig.~\ref{fig:brahms-pion} but now
for single-inclusive neutral pion production $pp\rightarrow \pi^0 X$
at mid-rapidities $|\eta|\le 0.35$ measured by  PHENIX \cite{ref:phenixpion}.
\label{fig:phenix-pion}}
\vspace*{-0.5cm}
\begin{center}
\vspace*{-0.6cm}
\epsfig{figure=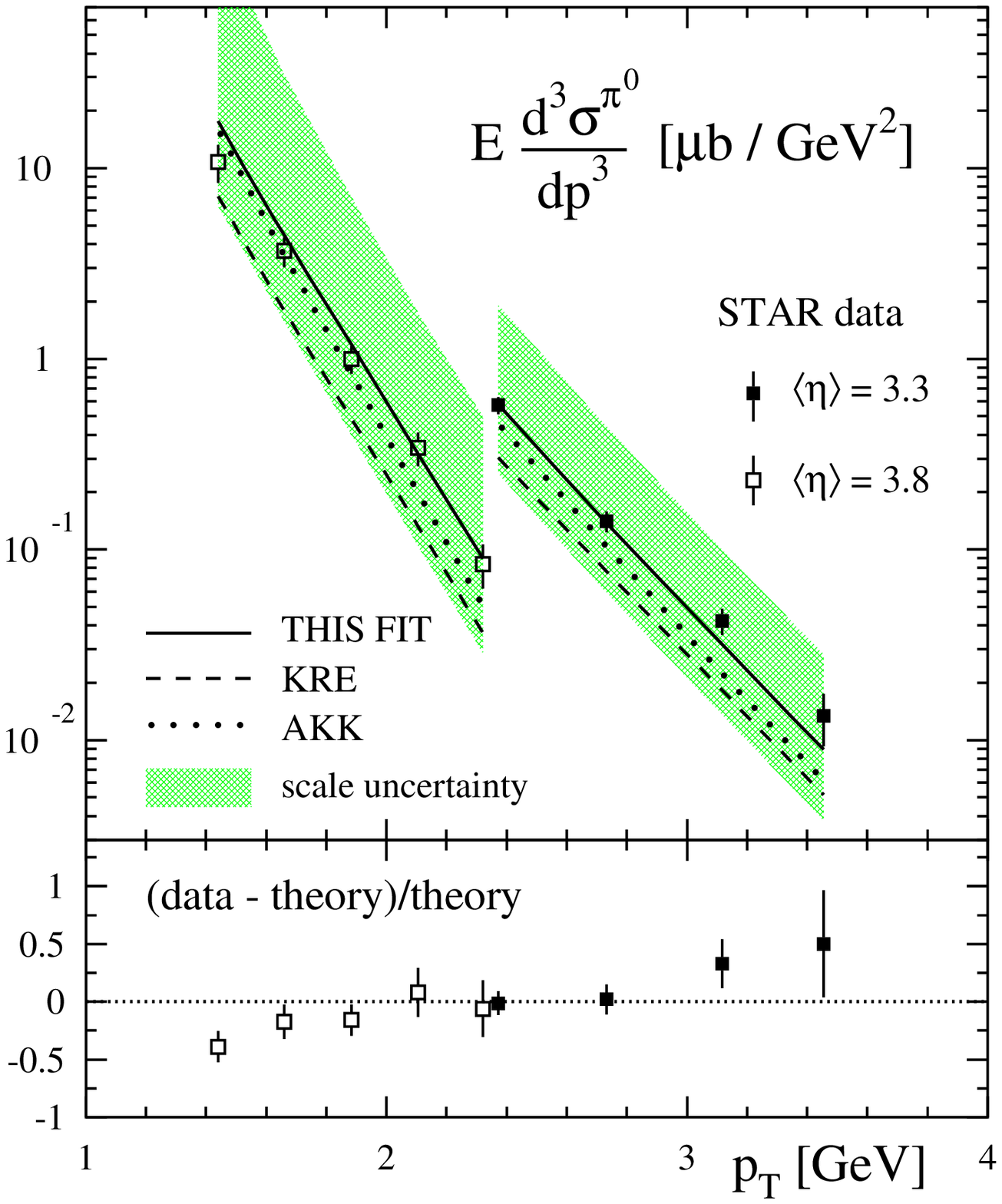,width=0.495\textwidth}
\end{center}
\vspace*{-0.7cm}
\caption{Same as in Fig.\ \ref{fig:phenix-pion} but now for
STAR data \cite{ref:starpion} at forward rapidities
$\langle\eta\rangle=3.3$ and $\langle\eta\rangle=3.8$,
solid and open squares, respectively.
\label{fig:star-pion}}
\vspace*{-0.5cm}
\end{figure}

\begin{figure}[th!]
\begin{center}
\vspace*{-0.6cm}
\epsfig{figure=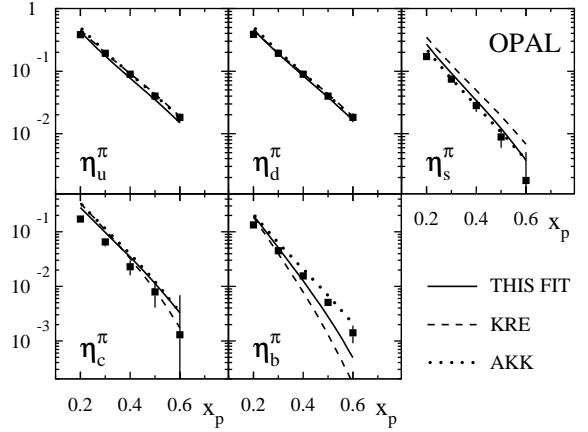,width=0.495\textwidth}
\end{center}
\vspace*{-0.7cm}
\caption{Comparison of the OPAL ``tagging probabilities''
\cite{ref:opaleta} for charged pions, $\eta_i^{\pi}$, as a
function of the minimum $x_p$,
see Eq.~(\ref{eq:opaleta}), with our NLO results (solid lines).
Also shown are the results obtained with the KRE \cite{ref:kretzer} and
AKK \cite{ref:akk} parameterizations, dashed and dotted lines, respectively.
\label{fig:opal-pion-eta}}
\vspace*{-0.5cm}
\end{figure}

%
From a first glance at Figures~\ref{fig:pioninclusive}-\ref{fig:opal-pion-eta},
one immediately notices the remarkable agreement between our new NLO fit
and data.
Experimental results for inclusive hadron production in SIA and
proton-proton collisions span several orders of magnitude, and the
energy scale of the different processes ranges from $1\,\text{GeV}$
to the mass of the $Z$-boson.
This strongly supports the underlying theoretical framework outlined
in Sec.~\ref{sec:framework}, in particular the fundamental notions of
factorization and universality for fragmentation functions.

%

Existing sets of NLO pion fragmentation functions \cite{ref:kretzer,ref:akk}
also give a nice overall description of the SIA data included in these
analyses, as indicated in Figs.~\ref{fig:pioninclusive}-\ref{fig:pion-cb}.
They fail, however, to satisfactorily reproduce charged pion production data
obtained in SIDIS and in proton-proton collisions,
Figs.~\ref{fig:sidis-pion} and \ref{fig:brahms-pion}, respectively.
In addition, estimates 
for neutral pion production rates in proton-proton collisions based
on KRE \cite{ref:kretzer} or AKK \cite{ref:akk}
fragmentation function differ substantially as can be seen in
Figs.~\ref{fig:phenix-pion} and \ref{fig:star-pion}.
On the contrary, our new set of NLO fragmentation functions
gives, for the first time, a nice {\em global} description of
hadron production data in electron-positron, lepton-nucleon, and
hadron-hadron scattering,
which constitutes a significant and necessary improvement.

The most significant difference between our NLO global analysis and previous
extractions of $D_i^{\pi}$ in \cite{ref:kretzer,ref:kkp,ref:akk,ref:hirai}
is the fact that we can now determine most aspects of the fragmentation
functions from data rather than being forced to make assumptions
due to the insufficient information contained in the SIA data alone.
We find that, in particular, the extra freedom regarding flavor symmetry
(or the lack thereof) as introduced in Eqs.~(\ref{eq:val_break}) and
(\ref{eq:sea_break}) allows us to reproduce the charged pion data
shown in Figs.~\ref{fig:sidis-pion} and \ref{fig:brahms-pion}.
In Table~\ref{tab:nlopionpara} we give the set of parameters
specifying the optimum fit of pion fragmentation functions at NLO accuracy
in Eq.~(\ref{eq:ff-input}) at our input scale $\mu_0=1\,\text{GeV}$ for the light flavors and the gluon,
and at $\mu_0=m_c=1.43\,\mathrm{GeV}$ and $\mu_0=m_b=4.3\,\mathrm{GeV}$,
for charm and bottom fragmentation, respectively.
As can be inferred from there, the outcome of the
global analysis deviates from the symmetry assumptions \cite{ref:kretzer,ref:hirai}
$N=1$ and $N^{\prime}=1$ in Eqs.(\ref{eq:val_break}) and (\ref{eq:sea_break})
by  more than 10\% and 20\%, respectively.

%
\begin{table}[t]
\caption{\label{tab:nlopionpara}Parameters describing the NLO
fragmentation functions for positively charged
pions, $D_i^{\pi^+}(z,\mu_0)$,
in Eq.~(\ref{eq:ff-input}) at the input scale $\mu_0=1\,\mathrm{GeV}$.
Inputs for the charm and bottom fragmentation functions refer to
$\mu_0=m_c=1.43\,\mathrm{GeV}$ and
 $\mu_0=m_b=4.3\,\mathrm{GeV}$, respectively.}
\begin{ruledtabular}
\begin{tabular}{cccccc}
flavor $i$ &$N_i$ & $\alpha_i$ & $\beta_i$ &$\gamma_i$ &$\delta_i$\\
\hline
$u+\overline{u}$ & 0.345&-0.015& 1.20&11.06& 4.23\\
$d+\overline{d}$ & 0.380&-0.015& 1.20&11.06& 4.23\\
$\overline{u}=d$ & 0.115& 0.520& 3.27&16.26& 8.46\\
$s+\overline{s}$ & 0.190& 0.520& 3.27&16.26& 8.46\\
$c+\overline{c}$ & 0.271&-0.905& 3.23& 0.00& 0.00\\
$b+\overline{b}$ & 0.501&-1.305& 5.67& 0.00& 0.00\\
$g$              & 0.279& 0.899& 1.57&20.00& 4.91\\
\end{tabular}
\end{ruledtabular}
\end{table}

Another crucial asset of our analysis is the enhanced
flexibility of the initial light quark and gluon
fragmentation functions as a function of $z$ in
Eq.~(\ref{eq:ff-input})
as compared to the standard three parameter
form used so far \cite{ref:kretzer,ref:kkp,ref:akk,ref:hirai}.
This is not only indispensable to accommodate SIDIS and
hadron-hadron scattering data but even somewhat improves the
quality of the fit to the SIA data.
Indeed, upon closer examination of Figs.~\ref{fig:pioninclusive}-\ref{fig:pion-cb},
in particular the ``(data-theory)/theory'' insets for each data set
on the right-hand side (r.h.s.) of each plot, one finds a slightly
improved overall agreement with data as compared to
the, still excellent, one for KRE and AKK; see, for instance,
the TPC or ALEPH data in Fig.~\ref{fig:pioninclusive}.
%
\begin{table}[ht]
\caption{\label{tab:exppiontab}Data used in the NLO global analysis
of pion fragmentation functions, the individual $\chi^2$ values for
each set, the fitted normalizations, and the total $\chi^2$ of the fit.}
\begin{ruledtabular}
\begin{tabular}{lcccc}
experiment& data & rel.\ norm.\ &data points & $\chi^2$ \\
          & type & in fit       & fitted     &         \\\hline
TPC \cite{ref:tpcdata}  & incl.\  &  0.94 & 17 & 18.5 \\
          & ``$uds$ tag''         &  0.94 &  9 & 1.9 \\
          & ``$c$ tag''           &  0.94 &  9 & 5.7 \\
          & ``$b$ tag''           &  0.94 &  9 & 7.4 \\
TASSO \cite{ref:tassodata}  &   incl. (34 GeV)      & 0.94      & 11 &  30.1    \\
          &   incl. (44 GeV)      &  0.94     &  7 &  20.5    \\
SLD \cite{ref:slddata}  & incl.\  &  1.008 & 28 & 14.0 \\
          & ``$uds$ tag''         &  1.008 & 17 & 11.6 \\
          & ``$c$ tag''           &  1.008 & 17 & 11.1  \\
          & ``$b$ tag''           &  1.008 & 17 & 33.2 \\
ALEPH \cite{ref:alephdata}    & incl.\  & 0.97 & 22 &  38.3 \\
DELPHI \cite{ref:delphidata}  & incl.\  & 1.0  & 17 & 42.3 \\
          & ``$uds$ tag''   &  1.0  & 17 & 26.4 \\
          & ``$b$ tag''     &  1.0  & 17 & 42.8 \\
OPAL \cite{ref:opaldata,ref:opaleta}  & incl.\ & 1.0 & 21 & 9.2 \\
          & ``$u$ tag'' &  1.10  & 5 & 11.8 \\
          & ``$d$ tag'' &  1.10  & 5 & 9.0  \\
          & ``$s$ tag'' &  1.10  & 5 & 49.8 \\
          & ``$c$ tag'' &  1.10  & 5 & 38.3 \\
          & ``$b$ tag'' &  1.10  & 5 & 73.0 \\\hline
HERMES \cite{ref:hermessidis}  & $\pi^+$ &  1.03 & 32 & 67.4 \\
                               & $\pi^-$ &  1.03 & 32 & 120.8 \\\hline
PHENIX \cite{ref:phenixpion}   & $\pi^0$ &  1.09 & 23 & 76.4 \\
STAR \cite{ref:starkaon}  & $\pi^0$, $\langle\eta\rangle=3.3$  & 1.05  & 4  & 3.4           \\
                          & $\pi^0$, $\langle\eta\rangle=3.7$  & 1.05  & 5  & 9.8           \\
BRAHMS \cite{ref:brahms}  & $\pi^+$,  $\langle\eta\rangle=2.95$  & 1.0 & 18 &  28.2             \\
                          & $\pi^-$, $\langle\eta\rangle=2.95$   & 1.0 & 18 &  43.0             \\ \hline\hline
{\bf TOTAL:} & & &392 & 843.7\\
\end{tabular}
\end{ruledtabular}
\end{table}

Noticeable is also that our fit follows the trend of the
data even below the $z$ values included in the analysis.
As for KRE \cite{ref:kretzer}, all data below $z_{\min}=0.05$
are not taken into account in the $\chi^2$ minimization
to ensure that the possible impact of small-$z$ resummations or hadron mass effects,
see Sec.~\ref{sec:ffs}, is negligible. In contrast, the agreement with data
rapidly deteriorates for AKK \cite{ref:akk} immediately below
$z=0.1$, from which they choose to exclude data from the fit.
This might be linked with the less flexible functional form
for the fragmentation functions.

In Tab.~\ref{tab:exppiontab} we list all data sets included in
our global analysis as discussed in Sec.~\ref{sec:datasets} and give
the individual $\chi^2$ values for each set. We note that
quoted $\chi^2$ values are based only on fitted data points, i.e.,
$z>0.05$ for SIA, and include normalization uncertainties
determined for each experiment in the fit.
Allowing for relative normalizations in a global analysis
within the range quoted by each experiment
is a common tool to ease possible tensions between certain
data sets. Indeed we find that the global fit considerably
improves after taking into account normalization ``shifts''.

In spite of the nice ``visual'' agreement between the fit and
data found in Figs.~\ref{fig:pioninclusive}-\ref{fig:opal-pion-eta},
the total $\chi^2$ of 843.7 units in Tab.~\ref{tab:exppiontab}
appears to be fairly large in view of the roughly 400 data points fitted.
For the SIA data, the large $\chi^2$ can be pinpointed to only very
few data points. Due to extremely high precision of the data on the 
$Z$-resonance, any deviation between data and theory is strongly penalized
in the $\chi^2$ evaluation and results in an overall $\chi^2$ per degree of freedom
which is rather large. This is a common ``characteristic'' of
all extractions of  fragmentation function made so far
\cite{ref:kretzer,ref:kkp,ref:akk,ref:hirai}. We also wish to
point out, that there is a tension between the behavior of the DELPHI
data at large $z$ and those of all the other data sets at $Q=M_Z$
which cannot be resolved by the fit,
see Figs.~\ref{fig:pioninclusive}-\ref{fig:pion-cb}.
The in general larger $\chi^2$ values of the heavy flavor, in particular
bottom quark, tagged SIA cross sections in Fig.~\ref{fig:pion-cb}
might be related to some extent to contaminations from weak decays.

We will further scrutinize the quality of the fit to
the SIDIS and hadronic data  in the following.
As illustrated in Fig.~\ref{fig:sidis-pion}, the agreement between the 
(preliminary) charged pion multiplicities in SIDIS from the HERMES 
experiment \cite{ref:hermessidis} and the results of our fit is remarkably 
good. The theoretical estimates for the multiplicities are computed using
PDFs from Ref.~\cite{ref:mrst} as input in Eq.~(\ref{eq:f1sidis}), 
although no significant differences are found using other 
modern sets, e.g., \cite{Pumplin:2005rh}. 

The fit not only
reproduces accurately the normalization in each $z$-bin for both
$\pi^+$ and $\pi^-$, but also the ``scaling violations'' in $Q^2$, which
are rather large at the low scales involved in the experiment.
Using the KRE fragmentation functions instead, also nicely reproduces the data in
the $z$ bins up to 0.6 for the $\pi^+$ multiplicities but considerably
overshoots all $\pi^-$ data, indicating that the {\em assumed}
favored/unfavored separation \cite{ref:kretzer}
$D_{\bar{u}}^{\pi^+}/D_u^{\pi^+}=(1-z)$ is not accurate enough.
We note that AKK \cite{ref:akk} refrains from imposing any 
assumptions in their fit, which are beyond what can be obtained
from SIA data. Hence, the AKK (and also KKP \cite{ref:kkp}) sets cannot be used
in theoretical calculations whenever the experiment discriminates
between positively or negatively charged pions (or kaons). This, of course,
seriously restricts the potential applications and testability of these sets.
%

Again, as can be inferred from Tab.~\ref{tab:exppiontab}, the
individual $\chi^2$ from SIDIS is fairly large, in particular for
the $\pi^-$ multiplicities. As in SIA, these data are based
on a sample with high statistics and hence the accuracy of the
data is very good. As before, any deviations between theory and
data are severely punished by a large contribution to $\chi^2$.
It is mainly the bin at the largest $z$ which makes all the trouble
and contributes most to the total $\chi^2$. The optimum fit
already ``negotiates'' the best compromise in describing these
data. As mentioned in Sec.~\ref{sec:datasets}, we
include an estimate of the theoretical uncertainties due
to finite bin-size effects in the $\chi^2$ minimization.
These are indicated by the shaded bands on the r.h.s.\
of Fig.~\ref{fig:sidis-pion}.
We also note, that we do not include the first data point
in each $z$ bin in the fit as the scale $Q$ almost coincides with
the already low input scale of our fit.
In general, one may wonder about possible contaminations from, say,
higher twists at the low $Q^2$ values accessible by HERMES.
However, we do not find any indications that
the SIDIS data are incompatible with the other data sets in our global
analysis.

In Figs.~\ref{fig:brahms-pion} - \ref{fig:star-pion} we compare
the results of our fit to recent data from proton-proton collisions
at $\sqrt{s}=200\,\text{GeV}$ at RHIC.
A general characteristics of all hadronic data is a large
theoretical uncertainty associated with the choice of the
arbitrary factorization and renormalization scales
$\mu_f$ and $\mu_r$, respectively, in Eq.~(\ref{eq:pp-xsec}).
Although largely reduced when going from the LO to the NLO approximation
for the $pp\to \pi X$ cross section, as demonstrated in
\cite{ref:jsv-pion,ref:aversa}, theoretical errors remain much more
sizable than experimental errors.
For our analysis we choose the transverse momentum of the produced pion
as the hard scale, i.e., $\mu_f=\mu_r=p_T$, which yields a very
good overall description at both central and forward rapidities.
The shaded bands in Figs.~\ref{fig:brahms-pion} - \ref{fig:star-pion}
indicate theoretical uncertainties when all scales are varied
simultaneously in the ``usual'' range $p_T/2\le \mu_f=\mu_r \le 2 p_T$.

In spite of the large uncertainties, the value of the RHIC data in
the global analysis is manifold: data at central rapidities $\eta\simeq 0$
and not too large $p_T$ are strongly dominated by $gg\to g X$ scattering
and hence constrain the gluon fragmentation $D_g^{\pi}$. At forward rapidities
$\eta\simeq 3$ the mixture between quark and gluon fragmentation
is roughly equal. It turns out that for both central and forward
rapidities, the fragmentation occurs at fairly large average
$\langle z \rangle \gtrsim 0.5$, see, e.g.,
Fig.~4 in Ref.~\cite{ref:guzey}, where information from SIA and SIDIS
is sparse.
As in case of SIDIS, the relevant hard scale of the process,
$Q={\cal{O}}(p_T)$, is much smaller than in SIA,
thereby allowing to exploit evolution effects to further
constrain the fragmentation functions.

The charged separated pion data obtained by BRAHMS very recently \cite{ref:brahms}
and shown in Fig.~\ref{fig:brahms-pion}, nicely back up the separation
of favored and unfavored fragmentation functions obtained from the
SIDIS data discussed above.
Another important feature of the BRAHMS and the RHIC data in
general, is the failure of the KRE set to reproduce them.
As can be seen in Figs.~\ref{fig:brahms-pion} - \ref{fig:star-pion},
using the KRE fragmentation one considerably undershoots all RHIC data. Only
by pushing the factorization scales to the extreme this could be
remedied to some extent. This observation has usually been taken as
an indication of an inadequately small gluon fragmentation function
in the KRE set at intermediate-to-large $z$ and scales of
a few GeV.
The fact that the agreement with the PHENIX data
in Fig.~\ref{fig:phenix-pion} is much better at
large $p_T$ when quark fragmentation becomes more
important, supports this picture.

\begin{figure}[th!]
\begin{center}
\vspace*{-0.6cm}
\epsfig{figure=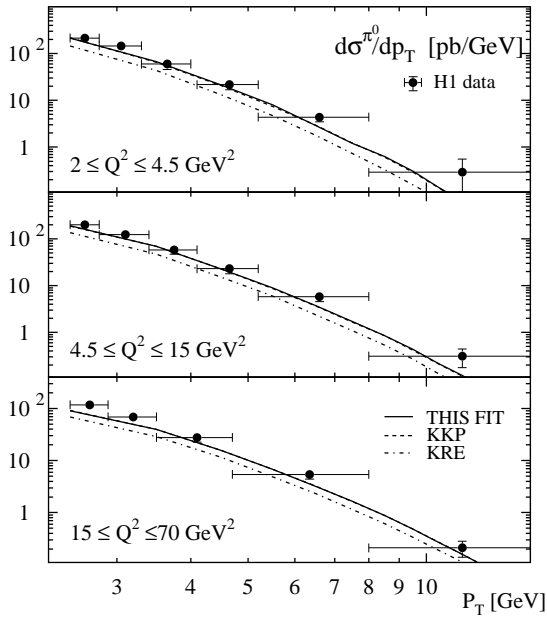,width=0.495\textwidth}
\end{center}
\vspace*{-0.7cm}
\caption{Transverse momentum distributions for neutral pions
obtained by H1 \cite{Aktas:2004rb} in deep-inelastic $e^+p$ collisions
compared to a NLO prediction using our new set of pion
fragmentation functions. Also shown are the results obtained with the
KRE \cite{ref:kretzer} and KKP \cite{ref:kkp} parameterizations.
\label{fig:HERA-plot}}
\vspace*{-0.5cm}
\end{figure}

The recent AKK set \cite{ref:akk} (as well as the preceding
KKP analysis \cite{ref:kkp}) are characterized by a much larger
gluon fragmentation function than in KRE and, consequently,
leads to a good description of the PHENIX data and, to a
lesser extent, also of the STAR data. The latter may
suggest the need for an even larger gluon fragmentation function.
As for SIDIS, the KKP or AKK sets cannot be used to compare
to the charge separated BRAHMS data.
Since the KRE, KKP, and AKK sets are based on the analysis of roughly
the same SIA data, the huge difference between the
obtained gluon distributions only demonstrates again that
a global analysis is imperative in obtaining reliable fragmentation
functions. We will discuss the individual fragmentation functions
and their uncertainties in some detail in Secs.~\ref{sec:ff-discussion}
and \ref{sec:unc-results}, respectively. We already note here, that,
surprisingly, the gluon fragmentation function obtained in our combined
fit turns out to be smaller than in AKK at intermediate $z$,
$0.2 \lesssim z \lesssim 0.5$, and only becomes larger at
$z\gtrsim 0.6$. This is indicative of the complex interplay of the
information provided by the different data sets in a global analysis.

The remaining data set used to constrain the pion fragmentation
functions are the OPAL ``tagging probabilities'' \cite{ref:opaleta}
as defined in Eq.~(\ref{eq:opaleta}).
As can be inferred from Fig.~\ref{fig:opal-pion-eta}, all
sets reproduce nicely the ``data'' for up and down flavors,
which indicates that they are already essentially fixed by
SIA data.
For strangeness fragmentation, however, the sets differ considerably.
In the KRE analysis \cite{ref:kretzer}, $D_s^{\pi}$ is completely
fixed by assuming $N^{\prime}=1$ in Eq.~(\ref{eq:sea_break}),
which overshoots significantly the OPAL result for $\eta_s^{\pi}$.
For heavy flavors, in particular for charm, the agreement is less
favorable for all sets. The results for $\eta_c^{\pi}$ from OPAL
are clearly at odds with the tagged SIA data from TPC and SLD
shown in Fig.~\ref{fig:pion-cb}. However,
as explained in Sec.~\ref{sec:datasets}, the interpretation of the
OPAL results beyond the LO should be taken with a grain of salt.
So is not surprising to find some discrepancy in our fit where, contrary
to AKK \cite{ref:akk}, the flavor separation comes not only from the
OPAL results but also from the interplay of other data sets, which have
a straightforward and reliable interpretation in pQCD.
The fairly large $\chi^2$ values we obtain for $\eta_s^{\pi}$,
$\eta_c^{\pi}$, and $\eta_b^{\pi}$ are therefore not alarming
and perhaps even expected.
Nevertheless, it is encouraging
that the general trend of the data is roughly reproduced by our
combined global fit. We have also checked, that the outcome
of our analysis does not change, if we excluded the
OPAL results from out fit.

Our newly obtained NLO pion fragmentation functions are best tested in predictions
for cross sections {\em not} included in the global analysis.
As a final cross check we therefore compare in Fig.~\ref{fig:HERA-plot}
measurements by the H1 collaboration \cite{Aktas:2004rb}
of forward neutral pions in deep-inelastic positron-proton collisions
at $\sqrt{s}\simeq 300\,\text{GeV}$ with NLO predictions based
on our new set of fragmentation functions.
The pions are required to be produced within a small angle $\theta_{\pi} \in[5^{\circ},25^o]$
from the direction of the proton beam in the laboratory frame, with an energy
fraction $x_{\pi}=E_{\pi}/E_P>0.01$ and transverse momenta in the range
$2.5<p_T<\,15\mbox{ GeV}$.
The observable has been shown to be crucially dependent on NLO contributions
associated to the gluon fragmentation function \cite{Daleo:2004pn}.
The nice agreement between the data and the NLO results
based on our new set of pion fragmentation functions is reassuring.

\subsection{LO analysis of pion fragmentation functions}
%
For completeness, we have also performed a global analysis
of the same set of data given in Tab.~\ref{tab:exppiontab} where
all observables, $\alpha_s$, and the scale evolution of
the fragmentation functions are computed at LO accuracy.
We use the same parametrization (\ref{eq:ff-input}) and
fitting procedure as in the NLO case and outlined in
Sec.~\ref{sec:analysis-outline}. The parameters
of the optimum LO fit are given in Tab.~\ref{tab:lopionpara}.

\begin{table}[b!]
\caption{\label{tab:lopionpara} As in Tab.~\ref{tab:nlopionpara}
but now describing the LO fragmentation functions for positively charged
pions, $D_i^{\pi^+}(z,\mu_0)$.}
\begin{ruledtabular}
\begin{tabular}{cccccc}
flavor $i$ &$N_i$ & $\alpha_i$ & $\beta_i$ &$\gamma_i$ &$\delta_i$\\
\hline
$u+\overline{u}$& 0.367&-0.228& 1.20& 5.29& 4.51\\
$d+\overline{d}$& 0.404&-0.228& 1.20& 5.29& 4.51\\
$\overline{u}=d$& 0.117& 0.123& 2.19& 7.80& 6.80\\
$s+\overline{s}$& 0.197& 0.123& 2.19& 7.80& 6.80\\
$c+\overline{c}$& 0.256&-0.310& 4.89& 0.00& 0.00\\
$b+\overline{b}$& 0.469&-1.108& 6.45& 0.00& 0.00\\
$g$             & 0.493& 1.179& 2.83&-1.00& 6.76\\
\end{tabular}
\end{ruledtabular}
\end{table}

\begin{table}[th!]
\caption{\label{tab:exppionlotab} Same as in Tab.~\ref{tab:exppiontab}
but now at LO accuracy.}
\begin{ruledtabular}
\begin{tabular}{lcccc}
experiment& data & rel.\ norm.\ &data points & $\chi^2$ \\
          & type & in fit       & fitted     &         \\\hline
TPC \cite{ref:tpcdata}  & incl.\  & 0.94  & 17 & 22.7 \\
          & ``$uds$ tag''         & 0.94  &  9 & 1.9 \\
          & ``$c$ tag''           & 0.94  &  9 & 5.6 \\
          & ``$b$ tag''           & 0.94  &  9 & 7.3 \\
TASSO \cite{ref:tassodata}  &   incl. (34 GeV)      & 0.94      & 11 &  48.1    \\
          &   incl. (44 GeV)      &  0.94     &  7 &  21.5    \\
SLD \cite{ref:slddata}  & incl.\  &  1.007 & 28 & 20.9 \\
          & ``$uds$ tag''         &  1.007 & 17 & 21.3 \\
          & ``$c$ tag''           &  1.007 & 17 & 9.3 \\
          & ``$b$ tag''           &  1.007 & 17 & 34.5 \\
ALEPH \cite{ref:alephdata}    & incl.\  & 0.97 & 22 &  64.4\\
DELPHI \cite{ref:delphidata}  & incl.\  & 1.0  & 17 & 45.9 \\
          & ``$uds$ tag''   &  1.0  & 17 & 30.6 \\
          & ``$b$ tag''     &  1.0  & 17 & 51.9\\
OPAL \cite{ref:opaldata,ref:opaleta}  & incl.\ & 1.0 & 21 & 20.7 \\
          & ``$u$ tag'' &  1.10  & 5 & 9.3\\
          & ``$d$ tag'' &  1.10  & 5 & 7.5\\
          & ``$s$ tag'' &  1.10  & 5 & 66.7\\
          & ``$c$ tag'' &  1.10  & 5 & 36.9\\
          & ``$b$ tag'' &  1.10  & 5 & 88.8\\\hline
HERMES \cite{ref:hermessidis}  & $\pi^+$ &  1.03 & 32 & 53.6 \\
                               & $\pi^-$ &  1.03 & 32 & 153.9 \\\hline
PHENIX \cite{ref:phenixpion}   & $\pi^0$ &  1.09 & 23 & 82.2 \\
STAR \cite{ref:starkaon}  & $\pi^0$, $\langle\eta\rangle=3.3$  & 0.95  & 4  &  15.5 \\
                          & $\pi^0$, $\langle\eta\rangle=3.7$  & 0.95  & 5  &  11.7 \\
BRAHMS \cite{ref:brahms}  & $\pi^+$,  $\langle\eta\rangle=2.95$  & 1.0 & 18 &  46.3 \\
                          & $\pi^-$, $\langle\eta\rangle=2.95$   & 1.0 & 18 &  77.7 \\ \hline\hline
{\bf TOTAL:} & & &392 & 1056.8\\
\end{tabular}
\end{ruledtabular}
\end{table}

As it can be immediately seen from Tab.~\ref{tab:exppionlotab}, the quality of the LO fit
is significantly worse than the NLO one, resulting in a 25\% increase in
the total $\chi^2$.
Although all individual observables show an increase in $\chi^2$, it is
somewhat more noticeable for proton-proton collision data.
This is a sensible and expected result as the NLO corrections are known
to be fairly large and important \cite{ref:jsv-pion,ref:aversa}.
To make up for the smaller LO partonic scattering cross sections relevant
for RHIC data, the most striking difference between the LO and NLO fragmentation
functions is found for gluons, while the moments for the
quark flavors remain rather stable; cf.\ Tabs.~\ref{tab:nlopionpara} and
\ref{tab:lopionpara}.
In spite of the larger $\chi^2$, for consistency, our LO sets should be used for
rough estimates of observables where NLO corrections are not yet available,
or in event generators based on matrix elements at LO accuracy.
Because of the limited usefulness of the LO set, we refrain from going into
any further details here.

\subsection{NLO analysis of kaon fragmentation functions}
%
Our NLO fits to single inclusive kaon production data
as compared to those for pions, reflect the sensible difference
in quality of both data sets. Even the most precise kaon production data
in SIA exhibit experimental uncertainties typically twice as large
as those found for pions.
%
\begin{table}[t!]
\caption{\label{tab:expkaontab}Data used in the NLO global analysis of
kaon fragmentation functions, the individual $\chi^2$ values for
each set, the fitted normalizations, and the total $\chi^2$ of the fit.}
\begin{ruledtabular}
\begin{tabular}{lcccc}
experiment& data & rel.\ norm.\ &data points & $\chi^2$ \\
          & type & in fit       & fitted     &         \\\hline
TPC \cite{ref:tpcdata}  & incl.\  &  0.94  & 12 & 9.5 \\
SLD \cite{ref:slddata}  & incl.\  &  0.983 & 18 & 14.4 \\
          & ``$uds$ tag''         &  0.983 & 10 & 14.4 \\
          & ``$c$ tag''           &  0.983 & 10 & 17.2 \\
          & ``$b$ tag''           &  0.983 & 10 & 15.2 \\
ALEPH \cite{ref:alephdata}    & incl.\  & 0.97 & 13 & 12.3 \\
DELPHI \cite{ref:delphidata}  & incl.\  & 1.0  & 12 & 1.0 \\
          & ``$uds$ tag''   &  1.0  & 12 & 2.3 \\
          & ``$b$ tag''     &  1.0  & 12 & 4.3 \\
OPAL \cite{ref:opaleta}  & ``$u$ tag'' &  1.10  & 5 & 6.5\\
          & ``$d$ tag'' &  1.10  & 5 & 9.9 \\
          & ``$s$ tag'' &  1.10  & 5 & 36.8 \\
          & ``$c$ tag'' &  1.10  & 5 & 44.9 \\
          & ``$b$ tag'' &  1.10  & 5 & 18.6 \\\hline
HERMES \cite{ref:hermessidis}  & $K^+$ &  1.03 & 24 & 15.0 \\
                               & $K^-$ &  1.03 & 24 & 79.3 \\\hline
STAR \cite{ref:starpion}  & $K^0_S$ & 0.95  & 14  &  40.0             \\
BRAHMS \cite{ref:brahms}  & $K^+$,  $\langle\eta\rangle=2.95$  & 1.0 & 18 & 28.8               \\
                          & $K^-$, $\langle\eta\rangle=2.95$   & 1.0 & 18 & 21.5              \\ \hline\hline
{\bf TOTAL:} & & & 232 & 394.1\\
\end{tabular}
\end{ruledtabular}
\end{table}
\begin{table}[h!]
\caption{\label{tab:nlokaonpara}Parameters describing the NLO
fragmentation functions for positively charged
kaons, $D_i^{K^+}(z,\mu_0)$, at the input scale $\mu_0=1\,\mathrm{GeV}$.
Inputs for the charm and bottom fragmentation functions refer to
$\mu_0=m_c=1.43\,\mathrm{GeV}$ and
 $\mu_0=m_b=4.3\,\mathrm{GeV}$, respectively.}
\begin{ruledtabular}
\begin{tabular}{cccccc}
flavor $i$ &$N_i$ & $\alpha_i$ & $\beta_i$ &$\gamma_i$ &$\delta_i$\\
\hline
$u+\overline{u}$           & 0.058& 0.705& 1.20&15.00& 6.02\\
$s+\overline{s}$           & 0.343&-0.065& 1.20& 4.36& 3.73\\
$d+\overline{d}$           & 0.016& 1.108&10.00&10.00& 3.28\\
$\overline{u}=s$           & 0.008& 1.108&10.00&10.00& 3.28\\
$c+\overline{c}$           & 0.196& 0.102& 4.56& 0.00& 0.00\\
$b+\overline{b}$           & 0.139&-0.584& 7.42& 0.00& 0.00\\
$g$                        & 0.017& 5.055& 1.20& 0.00& 0.00\\
\end{tabular}
\end{ruledtabular}
\end{table}
The potentially problematic low-$z$ region is expected to set in
earlier due to the larger kaon mass. To be on the safe side, we
raise the cut from $z_{\min}=0.05$ to $z_{\min}=0.1$.
%
\begin{figure*}[th!]
\begin{center}
\vspace*{-0.6cm}
\epsfig{figure=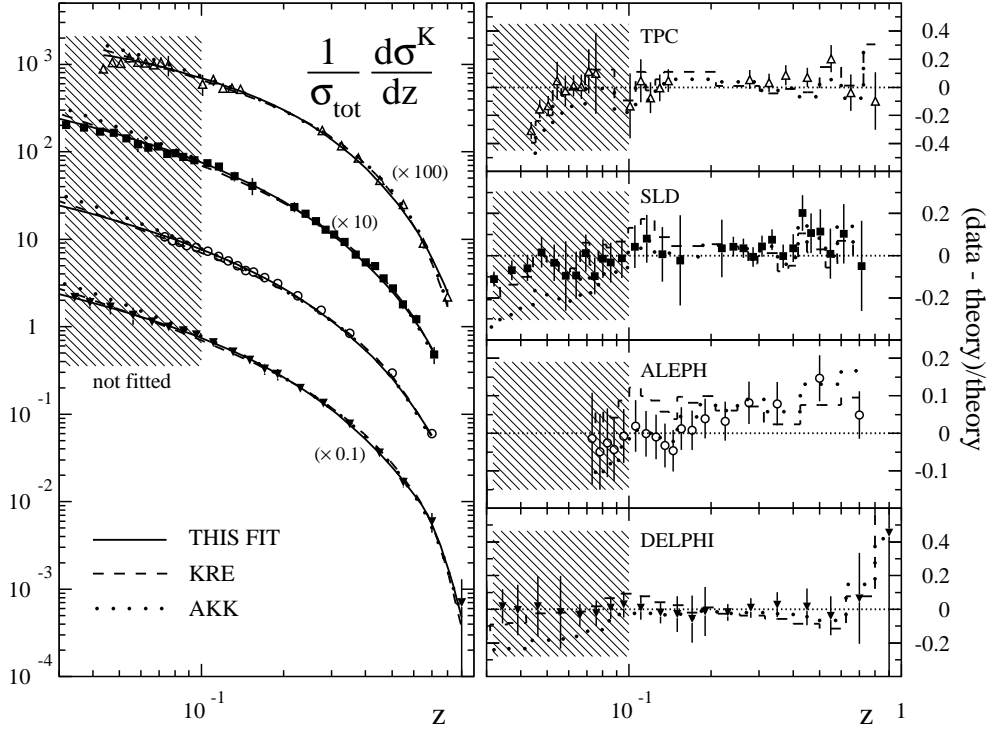,width=0.75\textwidth}
\end{center}
\vspace*{-0.7cm}
\caption{As in Fig.~\ref{fig:pioninclusive} but now for inclusive kaon
production in SIA.
\label{fig:kaoninclusive}}
\vspace*{-0.5cm}
\end{figure*}
\begin{figure}[t!]
\begin{center}
\vspace*{-0.6cm}
\epsfig{figure=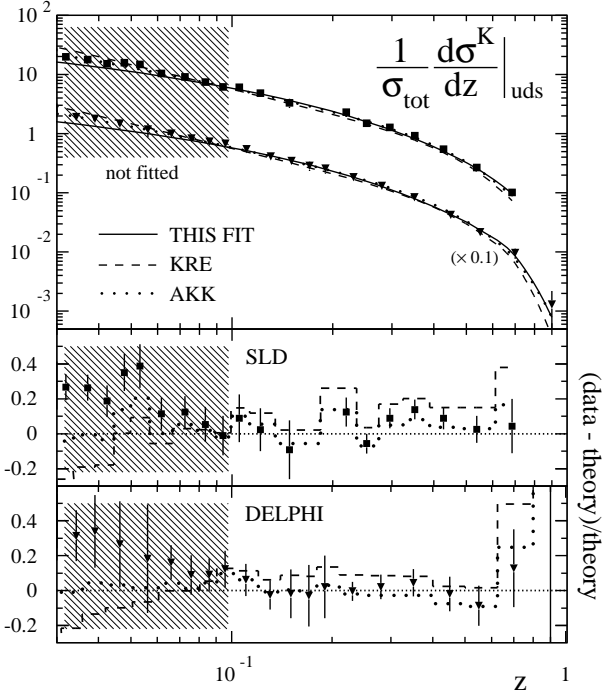,width=0.495\textwidth}
\end{center}
\vspace*{-0.7cm}
\caption{Same as in Fig.\ \ref{fig:kaoninclusive} but now for light quark (``uds'') tagged
cross sections. The upper panel shows the comparison of the ``uds'' tagged cross sections with data
and the lower two panels show ``(data-theory)/theory''.
\label{fig:kaon-uds}}
\vspace*{-0.5cm}
\end{figure}
Clearly, one must expect much
less well constrained fragmentation functions for kaons.
Otherwise the $\chi^2$ minimization proceeds along the same
lines as for the analysis of the pion data, and the results are
summarized in Tab.~\ref{tab:expkaontab}.
As demonstrated in Figs.~\ref{fig:kaoninclusive} - \ref{fig:opal-kaon-eta},
the overall agreement of our NLO fit and data is again remarkably good.
The set of parameters specifying the obtained kaon fragmentation functions
at NLO accuracy can be found in Tab.~\ref{tab:nlokaonpara}.

Starting with the SIA data in Figs.~\ref{fig:kaoninclusive} - \ref{fig:kaon-cb},
our fit shows again, thanks to the more flexible $z$-dependence of the initial
distributions (\ref{eq:ff-input}) and the less stringent flavor symmetry assumptions,
a slightly better agreement with SIA data than previous analyses
\cite{ref:kretzer,ref:akk}. Also, below $z_{\min}=0.1$ the fit still follows
the trend of the data, which ensures that finite mass effects
or the unstable small-$z$ behavior of the scale evolution
are still negligible.
In terms of $\chi^2$, none of the SIA data sets, even the ones for
heavy flavor tagged cross sections in Fig.~\ref{fig:kaon-cb},
poses any problems. The reduced statistical accuracy of the kaon data
compared to the pion data is immediately obvious and also reflects itself
in larger fluctuations in the ``(data-theory)/theory'' comparisons
in Figs.~\ref{fig:kaoninclusive} - \ref{fig:kaon-cb}.
\begin{figure*}[t!]
\begin{center}
\vspace*{-0.6cm}
\epsfig{figure=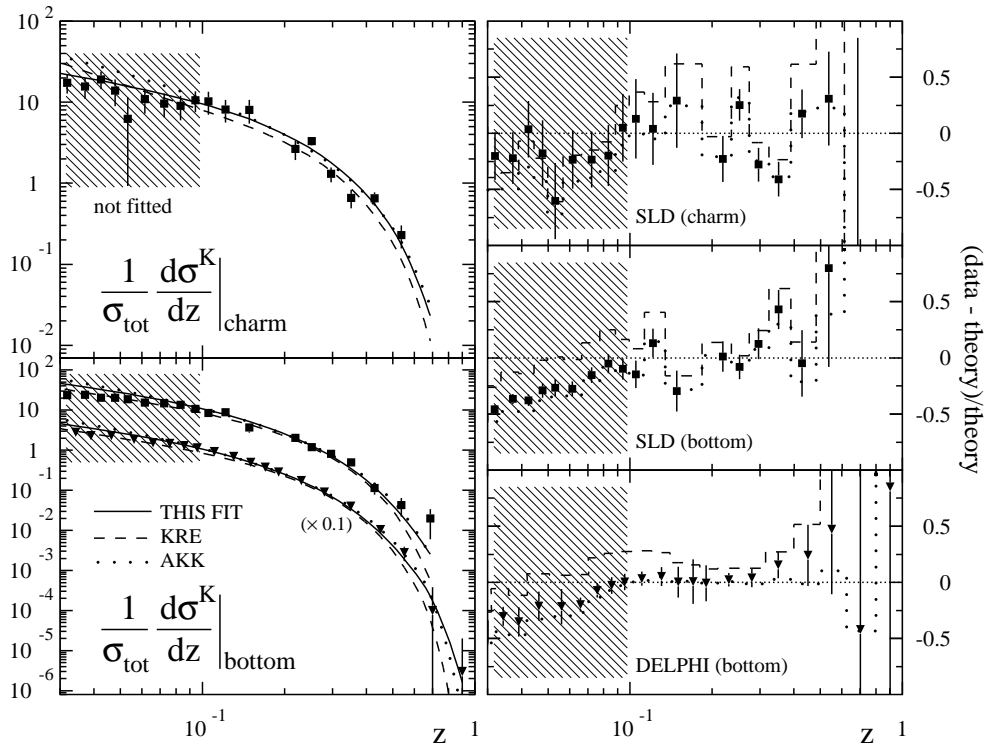,width=0.75\textwidth}
\end{center}
\vspace*{-0.7cm}
\caption{Same as in Fig.\ \ref{fig:kaoninclusive} but now for charm
and bottom quark tagged cross sections.
\label{fig:kaon-cb}}
\vspace*{-0.5cm}
\end{figure*}
\begin{figure*}[hb!]
\begin{center}
\vspace*{-0.6cm}
\epsfig{figure=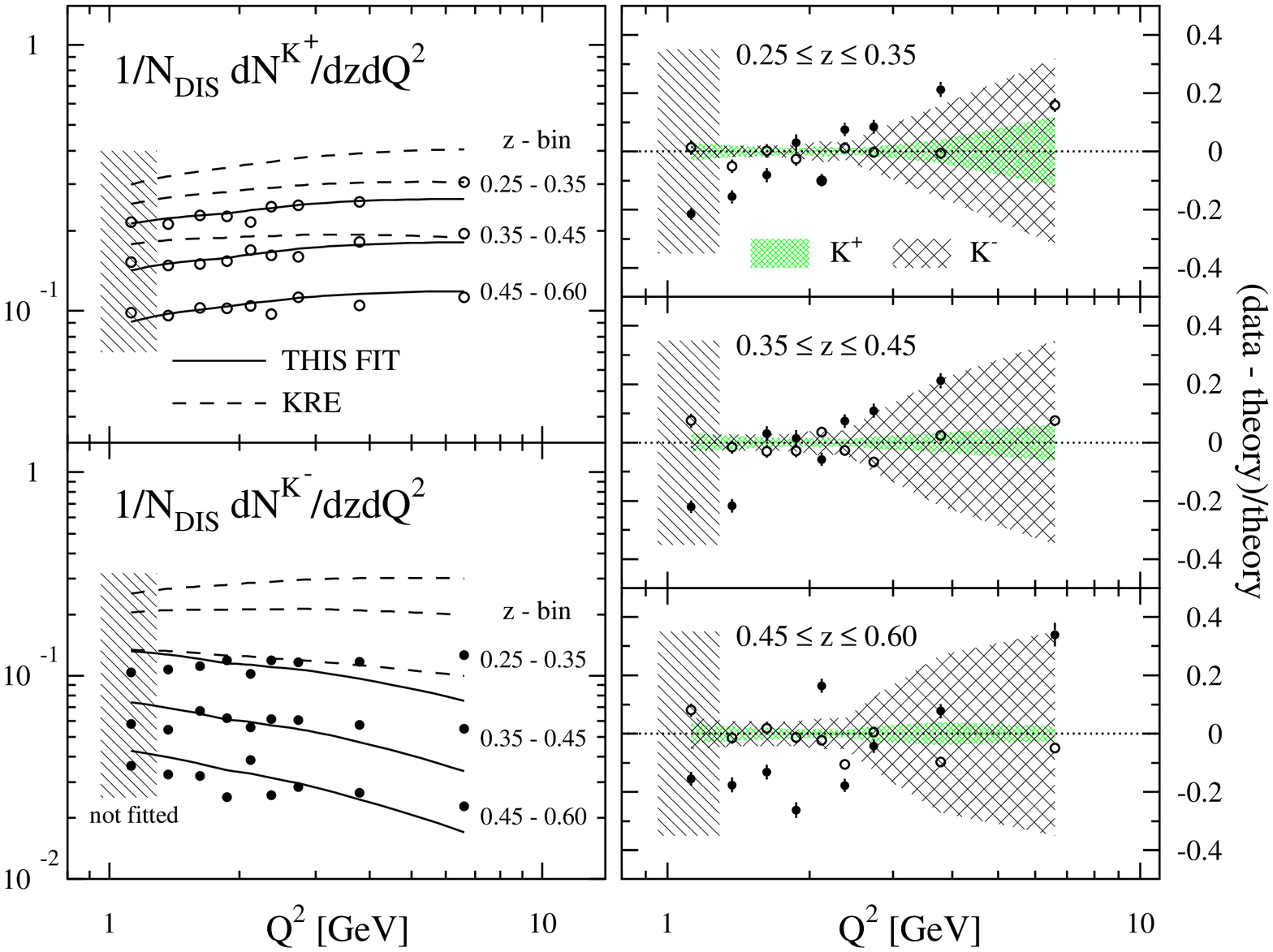,width=0.75\textwidth}
\end{center}
\vspace*{-0.7cm}
\caption{Same as in Fig.\ \ref{fig:sidis-pion} but now for
charged kaons.
\label{fig:sidis-kaon}}
\vspace*{-0.5cm}
\end{figure*}

The biggest asset of our global analysis is again that not only SIA data
but, for the first time, also SIDIS and RHIC data are nicely
reproduced as is demonstrated in Figs.~\ref{fig:sidis-kaon} - \ref{fig:star-kaon}.
It is worth noticing that, at variance with what happens for pions,
the HERMES SIDIS data, Fig.~\ref{fig:sidis-kaon},
rule out completely the flavor separation {\em assumed} in
Ref.~\cite{ref:kretzer} for kaon fragmentation functions. The prediction
based on the KRE set overshoots the data by a factor of two,
whereas our global fit shows much better agreement. This is in particular
true for the $K^+$ multiplicities, which in SIDIS receive the
dominant contribution related to the (large) up quark parton density 
in the proton. The production of $K^-$ is linked predominantly to 
strange quarks and anti-up quarks in the proton, both of which 
are much less abundant than up quarks for the relatively
large momentum fractions $x$ relevant for HERMES. 
Here our fit reproduces the magnitude of the $K^-$
multiplicities in each $z$ bin, but not the $Q^2$ slope. 
In fact, the related $\chi^2$ is the by far largest contribution 
to the total $\chi^2$ of the fit, cf.\ Tab.~\ref{tab:expkaontab}.
The theoretical uncertainty introduced by the size of the bins in
the computation of the cross section is also significantly larger than in the
case of pions. For negatively charged kaons it is further amplified by 
the rapid growth of sea quark densities in the proton with $Q^2$.
Taking also into account that the strange and, to some extent, 
also the anti-up quark parton densities are not too well constraint,
the agreement of the new fit is rather encouraging.

%
\begin{figure*}[t!]
\begin{center}
\vspace*{-0.6cm}
\epsfig{figure=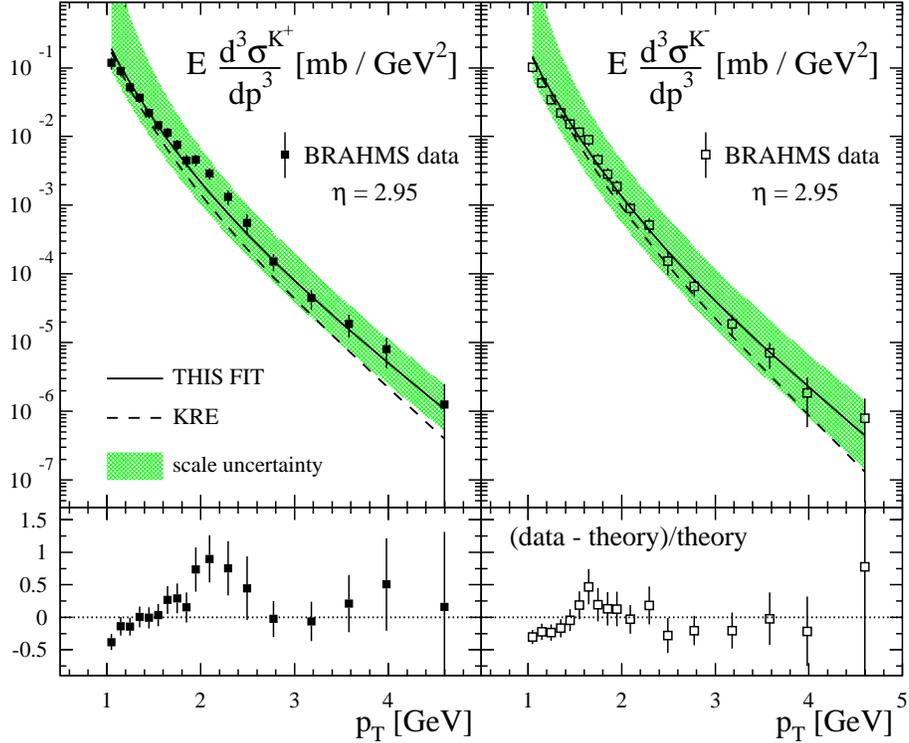,width=0.75\textwidth}
\end{center}
\vspace*{-0.7cm}
\caption{Same as in Fig.\ \ref{fig:brahms-pion} but now for
charged kaons.
\label{fig:brahms-kaon}}
\vspace*{-0.5cm}
\end{figure*}
\begin{figure}
\begin{center}
\vspace*{-0.6cm}
\epsfig{figure=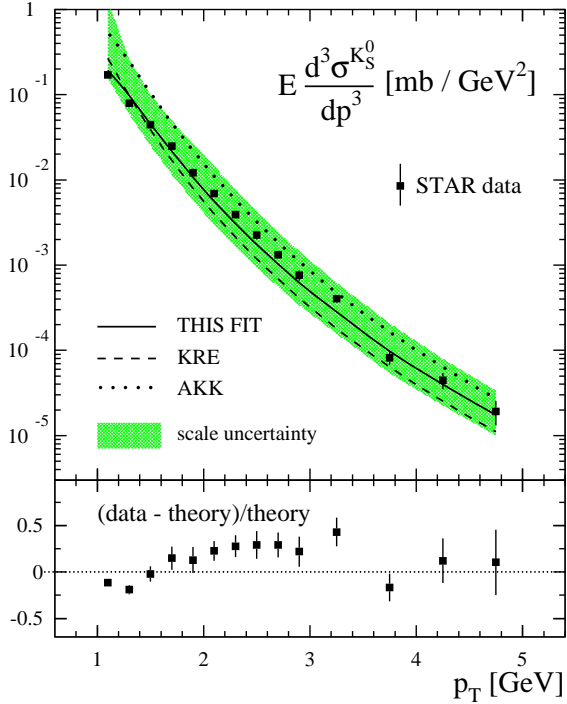,width=0.495\textwidth}
\end{center}
\vspace*{-0.7cm}
\caption{Same as in Fig.\ \ref{fig:phenix-pion} but now for
STAR data \cite{ref:starkaon} for $K^0_{S}$ production
at central rapidities $|\eta|\le 0.5$.
\label{fig:star-kaon}}
\vspace*{-0.5cm}
\end{figure}
\begin{figure}
\begin{center}
\vspace*{-0.6cm}
\epsfig{figure=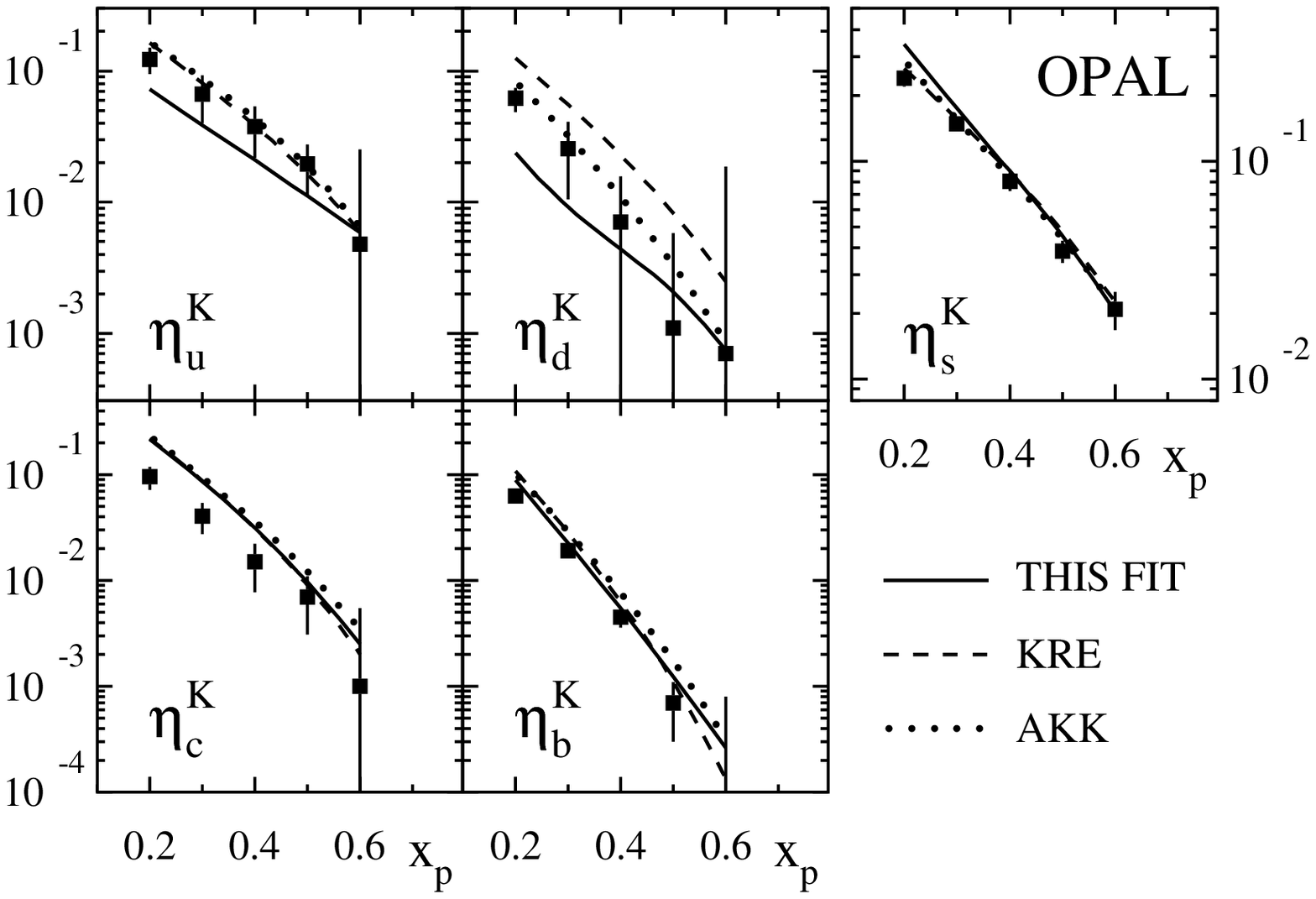,width=0.495\textwidth}
\end{center}
\vspace*{-0.7cm}
\caption{Same as in Fig.\ \ref{fig:opal-pion-eta} but now for
charged kaons.
\label{fig:opal-kaon-eta}}
\vspace*{-0.5cm}
\end{figure}
Our new set of kaon fragmentation functions also yields the best description of
the BRAHMS \cite{ref:brahms} and STAR data \cite{ref:starkaon},
Figs.~\ref{fig:brahms-kaon} and \ref{fig:star-kaon}, respectively,
which both probe the large $z$ regime, $z\gtrsim 0.5$, poorly
mapped out by SIA data.
As for pions, the KRE set undershoots all RHIC kaon data, whereas AKK
overshoots the STAR $K^0_S$ data. As before, the AKK (and KKP) set does not
allow to compute the charge separated $K^+$ and $K^-$ yields by BRAHMS.
Again, the discrepancy with the RHIC data can be traced back to the 
behavior of the gluon fragmentation function at large $z$ and will be
discussed further in Sec.~\ref{sec:ff-discussion}.
Notice again the very large theoretical scale uncertainties for the RHIC data.

The last data set entering the global analysis of kaon fragmentation
functions are the OPAL tagging probabilities $\eta_i^K$ \cite{ref:opaleta}.
In view of the already discussed conceptual problems with these data,
the overall agreement with the OPAL results in Fig.~\ref{fig:opal-kaon-eta}
is reasonable. The lack of agreement, most noticeable for charm,
again suggests some degree of inconsistency with other data sets included
in the fit.

\subsection{LO analysis of kaon fragmentation functions}

The LO global analysis of kaon production data yields 
significantly larger values of $\chi^2$. The parameters describing the 
optimum LO kaon fragmentation functions are given in Tab.~\ref{tab:lokaonpara} 
 while $\chi^2$ contributions are in Tab.~\ref{tab:loexpkaontab}.

\begin{table}[b!]
\caption{\label{tab:lokaonpara} As in Tab.~\ref{tab:nlokaonpara} but now for
the LO fragmentation functions for positively charged kaons.}
\begin{ruledtabular}
\begin{tabular}{cccccc}
flavor $i$ &$N_i$ & $\alpha_i$ & $\beta_i$ &$\gamma_i$ &$\delta_i$\\
\hline
$u+\overline{u}$& 0.054& 1.018& 1.20&15.00& 6.04\\
$s+\overline{s}$& 0.361& 0.733& 1.20&20.00& 5.28\\
$\overline{u}=s$& 0.005& 1.322&10.00&10.00& 3.67\\
$d+\overline{d}$& 0.010& 1.322&10.00&10.00& 3.67\\
$c+\overline{c}$& 0.214& 0.239& 4.27& 0.00& 0.00\\
$b+\overline{b}$& 0.147&-0.464& 7.37& 0.00& 0.00\\
$g$             & 0.036& 5.282& 1.20& 0.00& 0.00\\
\end{tabular}
\end{ruledtabular}
\end{table}

The overall increase with respect to the NLO fit is about 30\%, even larger
than the one found in the case of pions, and with the most noticeable
differences in the partial contributions stemming from RHIC and SIDIS data.


\begin{table}[t!]
\caption{\label{tab:loexpkaontab} Same as in Tab.~\ref{tab:expkaontab}
but now at LO accuracy.}
\begin{ruledtabular}
\begin{tabular}{lcccc}
experiment& data & rel.\ norm.\ &data points & $\chi^2$ \\
          & type & in fit       & fitted     &         \\\hline
TPC \cite{ref:tpcdata}  & incl.\  &  0.94  & 12 & 12.7 \\
TASSO \cite{ref:tassodata}  &   incl. (34 GeV)      & 0.94      & 4 &  2.6    \\
SLD \cite{ref:slddata}  & incl.\  &  0.983 & 18 & 18.1 \\
          & ``$uds$ tag''         &  0.983 & 10 & 21.9 \\
          & ``$c$ tag''           &  0.983 & 10 & 18.4 \\
          & ``$b$ tag''           &  0.983 & 10 & 15.0 \\
ALEPH \cite{ref:alephdata}    & incl.\  & 0.97 & 13 & 14.0 \\
DELPHI \cite{ref:delphidata}  & incl.\  & 1.0  & 12 & 1.2 \\
          & ``$uds$ tag''   &  1.0  & 12 & 2.6 \\
          & ``$b$ tag''     &  1.0  & 12 & 4.1 \\
OPAL \cite{ref:opaleta}  & ``$u$ tag'' &  1.10  & 5 & 7.8\\
          & ``$d$ tag'' &  1.10  & 5 & 10.6 \\
          & ``$s$ tag'' &  1.10  & 5 & 32.9 \\
          & ``$c$ tag'' &  1.10  & 5 & 53.1 \\
          & ``$b$ tag'' &  1.10  & 5 & 19.7 \\\hline
HERMES \cite{ref:hermessidis}  & $K^+$ &  1.03 & 24 & 23.9 \\
                               & $K^-$ &  1.03 & 24 & 131.2 \\\hline
STAR \cite{ref:starpion}  & $K^0_S$ & 0.95  & 14  &  59.0             \\
BRAHMS \cite{ref:brahms}  & $K^+$,  $\langle\eta\rangle=2.95$  & 1.0 & 18 & 36.9               \\
                          & $K^-$, $\langle\eta\rangle=2.95$   & 1.0 & 18 & 34.7              \\ \hline\hline
{\bf TOTAL:} & & & 236 & 520.7\\
\end{tabular}
\end{ruledtabular}
\end{table}

\subsection{\label{sec:ff-discussion}Fragmentation Functions}
%
\begin{figure*}
\begin{center}
\vspace*{-0.6cm}
\epsfig{figure=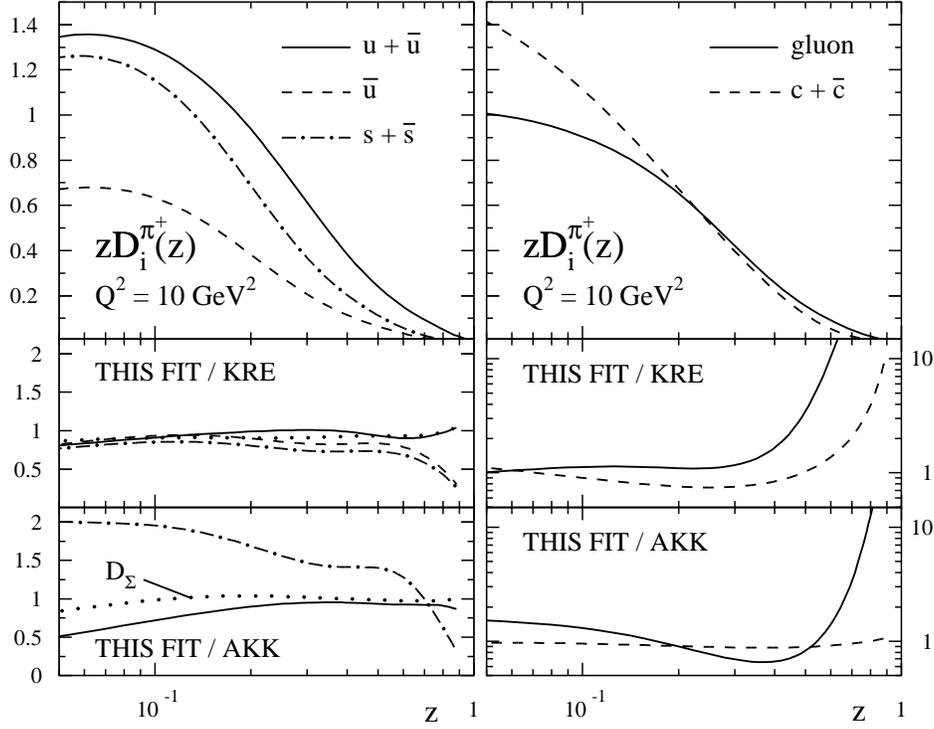,width=0.75\textwidth}
\end{center}
\vspace*{-0.7cm}
\caption{{\bf upper panels} individual fragmentation functions for positively
charged pions $zD_i^{\pi^{+}}(z,Q^2)$ at $Q^2=10\,\mathrm{GeV}^2$
for $i=u+\bar{u},\, \bar{u},\, s+\bar{s},\, g,$ and $c + \bar{c}$.
{\bf middle panels} ratios of our fragmentation functions to the ones
of KRE \cite{ref:kretzer}. The dotted line indicates the ratio for singlet
combination of fragmentation functions $zD_{\Sigma}^{\pi^{+}}$.
{\bf lower panels} ratios of our fragmentation functions to the ones
of AKK \cite{ref:akk}; note that $D_{\bar{u}}^{\pi^{+}}$ is not
available in the AKK analysis.
\label{fig:ff-pion-10}}
\vspace*{-0.5cm}
\end{figure*}
\begin{figure*}
\begin{center}
\vspace*{-0.6cm}
\epsfig{figure=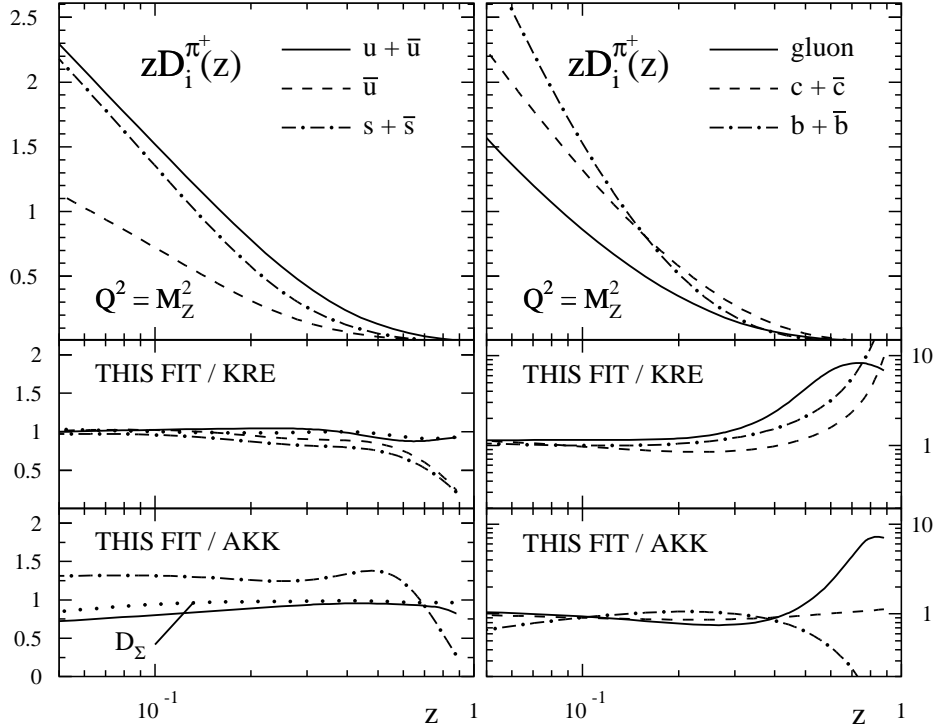,width=0.75\textwidth}
\end{center}
\vspace*{-0.7cm}
\caption{Same as in Fig.\ \ref{fig:ff-pion-10} but now for $Q = M_Z$.
Also shown is the $b+\bar{b}$ fragmentation function.
\label{fig:ff-pion-mz}}
\vspace*{-0.5cm}
\end{figure*}
\begin{figure*}
\begin{center}
\vspace*{-0.6cm}
\epsfig{figure=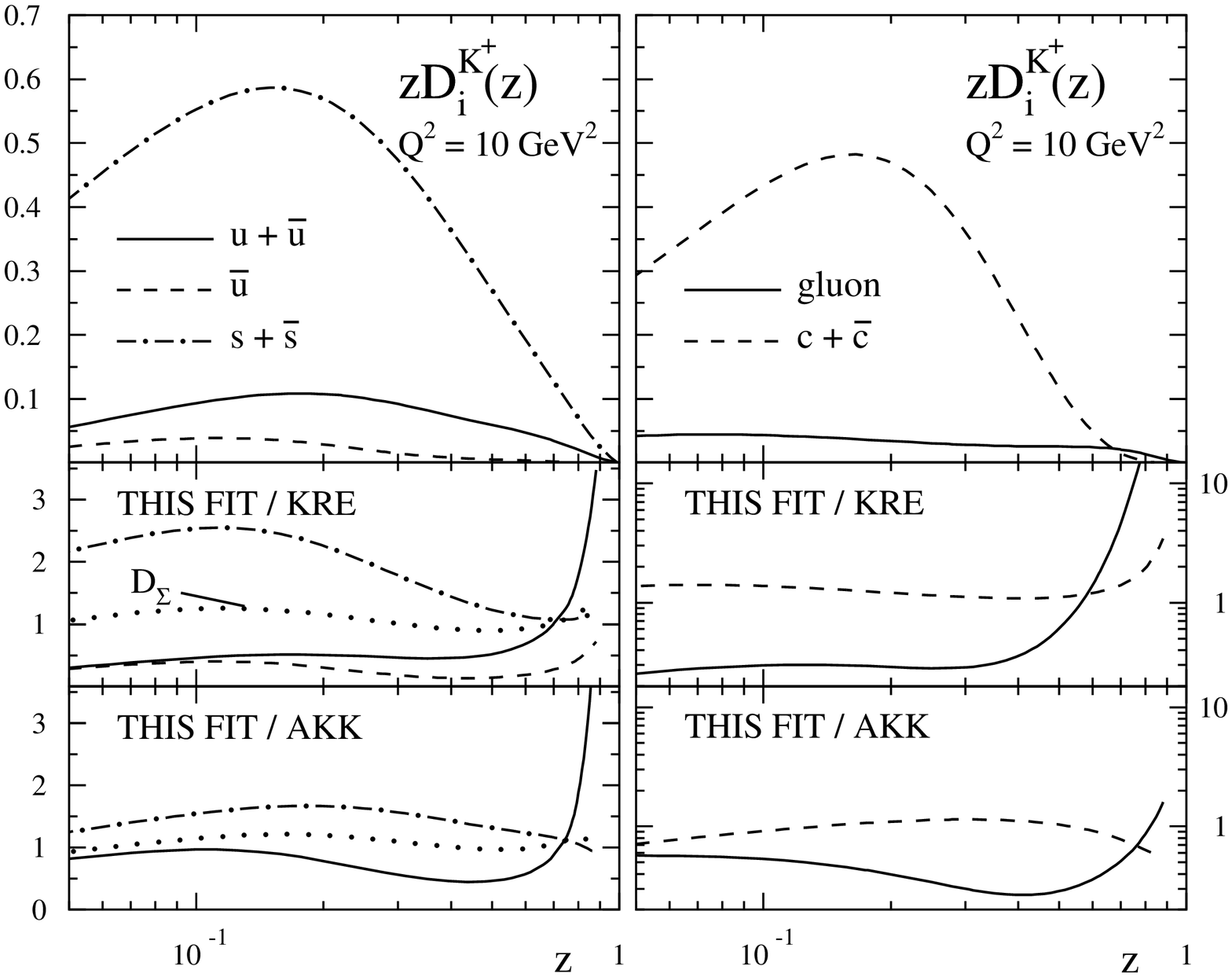,width=0.75\textwidth}
\end{center}
\vspace*{-0.7cm}
\caption{Same as in Fig.\ \ref{fig:ff-pion-10} but now for positively
charged kaons $K^+$.
\label{fig:ff-kaon-10}}
\vspace*{-0.5cm}
\end{figure*}
\begin{figure*}
\begin{center}
\vspace*{-0.6cm}
\epsfig{figure=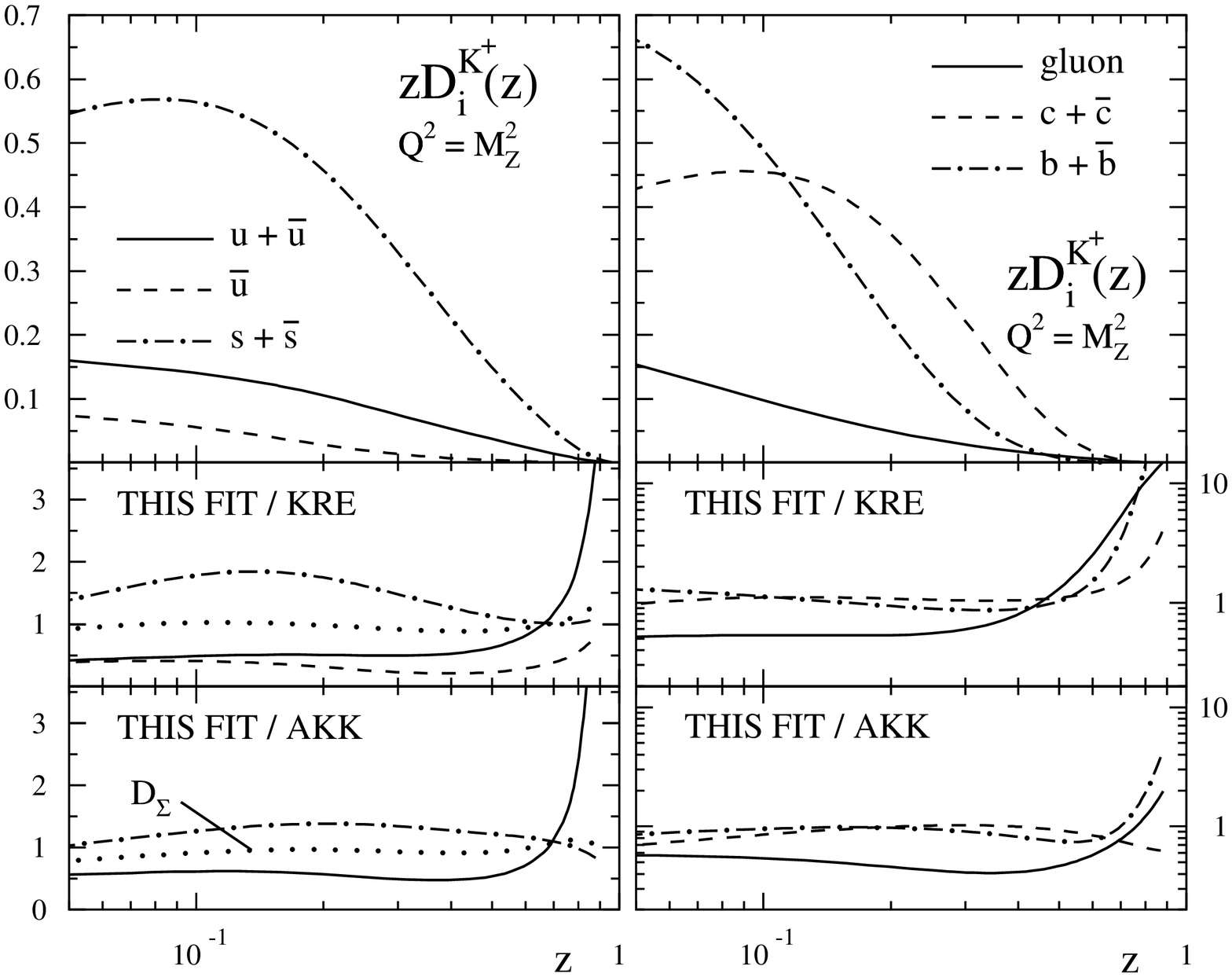,width=0.75\textwidth}
\end{center}
\vspace*{-0.7cm}
\caption{Same as in Fig.\ \ref{fig:ff-pion-mz} but now for positively
charged kaons $K^+$.
\label{fig:ff-kaon-mz}}
\vspace*{-0.5cm}
\end{figure*}

In this Section we shall present an overall description of the different
fragmentation functions $D_i^{\pi^+,K^+}$ obtained in the global 
fit \footnote{A {\tt Fortran} package containing our LO and NLO sets of 
pion and kaon fragmentation functions can be obtained upon request from the authors.}
and perform a
comparison with the KRE \cite{ref:kretzer} and AKK \cite{ref:akk} 
NLO sets based only on SIA data. 

The upper panels of Fig.~\ref{fig:ff-pion-10} show the fragmentation functions for
positively charged pions at the scale $Q^2=10$ GeV$^2$. As expected,
the sum $u+\bar{u}$ dominates over the {\it unfavored} distributions
$\bar{u}$ and $s$. At large values of $z$, there is an important
contribution from the valence $u$ fragmentation, while at small $z$
the sea distribution dominates and $u+\bar{u}\simeq 2 \bar{u}$. In the
same limit, it can be observed that the $s$ fragmentation function
turns out to be smaller than the corresponding $\bar{u}$ sea distribution,
as anticipated in Sec.~\ref{sec:nlopion} when discussing the value of the
relevant parameter $N'$. 
As can also be noticed from Fig.~\ref{fig:ff-pion-10}, charm and gluon 
fragmentation are quite sizable and comparable to the one of
the light quarks at small $z$. 
This is actually a general feature of heavy quark
fragmentation, opposite to the behavior of the usually less relevant
heavy quark parton distributions. At this scale, the bottom channel
has not opened yet, but the corresponding distribution can be
observed in Fig.~\ref{fig:ff-pion-mz}, where the same functions 
are plotted at a higher scale $Q^2=M_Z^2$. 
As expected, heavy quark and gluon fragmentation
are rather suppressed at larger values of $z$. 

In the middle and lower panels of Figs.~\ref{fig:ff-pion-10} and \ref{fig:ff-pion-mz},
we compare our set of fragmentation functions to those from KRE and
AKK, respectively. The largest differences appear for the {\it
unfavored} quark and gluon distributions and, usually, both at large $z$ 
and near $z_{\min}$ below which fragmentation functions cannot be used.
Notice that for AKK, $D_{\bar{u}}^{\pi^+}$ is not available for comparison 
and that their analysis is limited to $z>0.1$ rather than $z>z_{\min}=0.05$. 
Since AKK tends to overestimate the SIA cross-section outside the
fitted region, i.e., below $z=0.1$, any disagreement there is not
surprising.

While the discrepancy with KRE for the light quark distributions are
reasonably moderate, there happens to be a rather large difference
at the level of the strange fragmentation with AKK. The origin of this
disagreement can be easily understood: SIDIS data, not included in the
AKK fit, require a smaller fragmentation for $u$ (as can be seen in
the same plot) and $d$ quarks. Since the singlet combination,
$D_\Sigma$, is already well constraint by SIA data, as can also be seen
in Figs.~\ref{fig:ff-pion-10} and \ref{fig:ff-pion-mz},
that automatically requires an increase in the $D_s^{\pi^+}$ fragmentation
function.

The most pronounced differences appear at very large values of $z$ for 
both heavy flavor and gluon fragmentation functions.
In the heavy quark case, both SIDIS and RHIC data are insensitive and
SIA data suffer from very large experimental uncertainties for $z>0.4$
and even conflicting results. Therefore, in that kinematical regime
those densities cannot be determined well, and large discrepancies 
between different analyses are expected.
For the gluon fragmentation, the main source of information at $z\gtrsim 0.5 $
stems from PHENIX data at mid and STAR and BRAHMS data at forward rapidity.
These are included only in the analysis presented here. 
BRAHMS and STAR data access the largest $z$ values of all data sets in
our global analysis. 
As was observed in Figs.~\ref{fig:brahms-pion} - \ref{fig:star-pion},
the KRE sets underestimates all experimental data (for $\mu_f=\mu_r=p_T$),
while the AKK set agrees with the PHENIX data but tends to underestimate STAR.
This explains the differences in Figs.~\ref{fig:ff-pion-10} and \ref{fig:ff-pion-mz}
between our gluon fragmentation function and those of KRE and AKK. 
RHIC data tend to favor a significantly larger $D_g^{\pi^+}$ at large $z$.

Finally, the comparison between the different distributions and ratios in
Figs.~\ref{fig:ff-pion-10} and \ref{fig:ff-pion-mz} also highlights
the importance of the $Q^2$ evolution. Even though the distributions
show a much stronger rise at small $z$ when the scale is increased
\footnote{Recall that, as discussed in Sec.~\ref{sec:ffs}, 
at much smaller values of $z$ the NLO splitting kernels produce 
negative distributions and hadron mass effects become important.},
one also finds a much better agreement between the different sets 
at $Q^2=M_Z^2$, where most of the very precise SIA data are obtained. 
In other words, the evolution downwards from $M_Z$, were the
distributions agree best, to scales relevant for RHIC and SIDIS
data, exacerbate the differences between them.

Figures \ref{fig:ff-kaon-10} and \ref{fig:ff-kaon-mz} provide the
same information and comparisons as in the previous ones but for
positively charged kaons. As expected, the dominant fragmentation in
the light quark sector corresponds to the strange distribution.
While heavy quark densities are as large as those for pions, the
gluon fragmentation turns out to be much less sizable, even though it is
still larger than those from KRE and AKK at large $z$ in order to
fit the proton-proton data from STAR and, in particular,
at forward rapidities from BRAHMS. 
The comparison in the light quark sector shows many similarities with
the pion case, but here the discrepancies are more noticeable with
the KRE set instead. The global fit requires a smaller contribution
from $u$ quark fragmentation, mostly from SIDIS data (distributions 
like KRE overestimate SIDIS data, see Fig.~\ref{fig:sidis-kaon}) 
resulting in an increase in the strange sector as the singlet
$D_{\Sigma}^{K^+}$ is again constrained by SIA data.

\subsection{\label{sec:unc-results} Uncertainties}

\begin{figure*}[th!]
\begin{center}
\vspace*{-.6cm}
\epsfig{figure=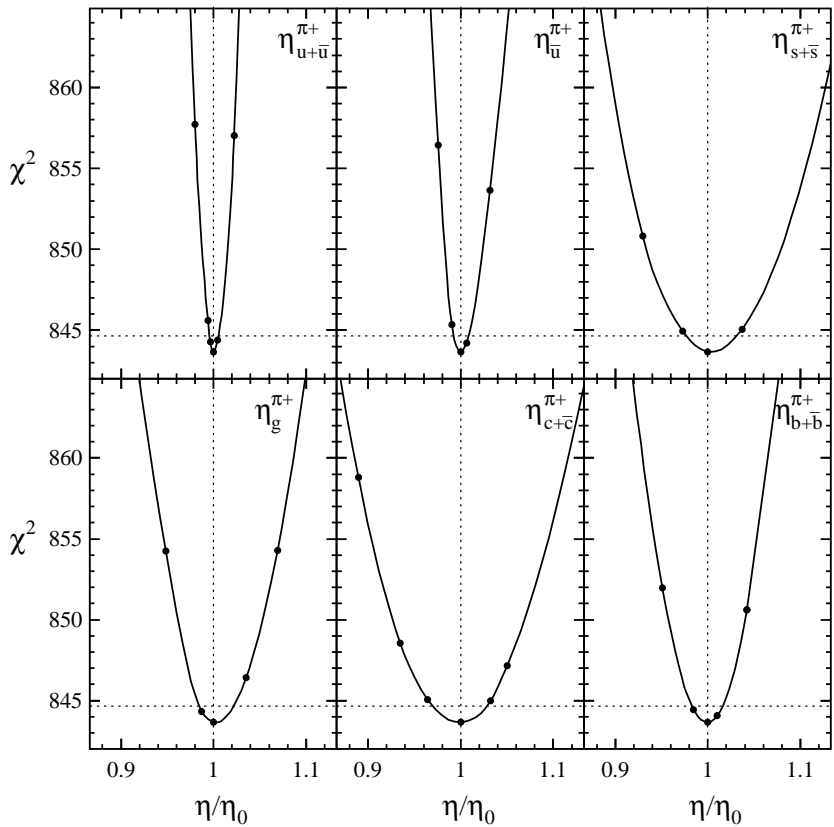,width=0.75\textwidth}
\end{center}
\vspace*{-.7cm}
\caption{Profiles of $\chi^2$ for the NLO pion fragmentation fit
as a function of the truncated second moments
$\eta^{\pi^+}_i(x_p=0.2,Q=5\,\text{GeV})$ for different flavors. 
The moments are
normalized to the value $\eta^{\pi^+}_{i\,0}$ they take in the best fit
to data.
\label{fig:pion-uncert}}
\vspace*{-0.5cm}
\end{figure*}
\begin{figure*}
\begin{center}
\vspace*{-.6cm}
\epsfig{figure=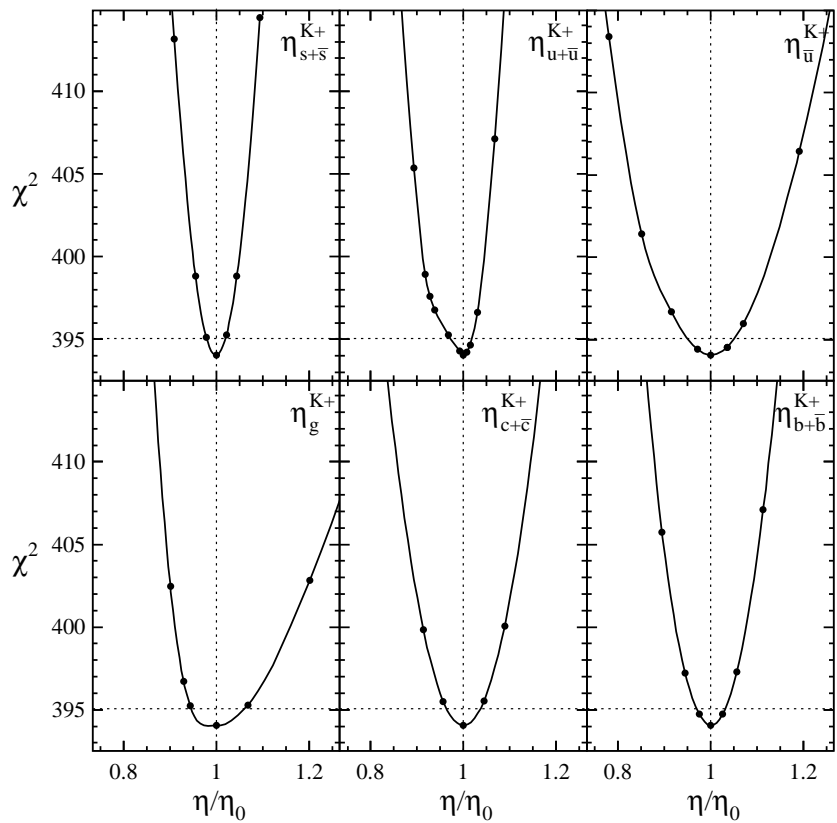,width=0.75\textwidth}
\end{center}
\vspace*{-.7cm}
\caption{The same as in Fig.~\ref{fig:pion-uncert}, but now for kaon fragmentation
at NLO.
\label{fig:kaon-uncert}}
\vspace*{-0.5cm}
\end{figure*}

\begin{figure}
\begin{center}
\vspace*{-0.6cm}
\epsfig{figure=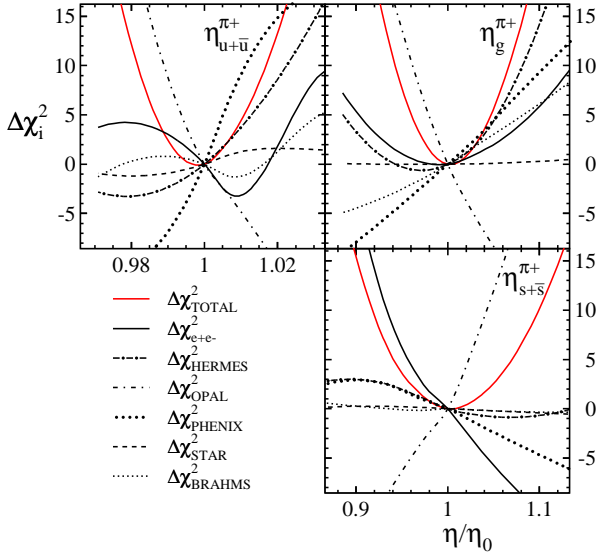,width=0.495\textwidth}
\end{center}
\vspace*{-0.7cm}
\caption{Partial contributions $\Delta \chi^2_i$ 
of the individual data sets to the NLO global analysis of pion fragmentation 
functions against the variation in selected truncated moments.
\label{fig:pion-partial}}
\vspace*{-0.5cm}
\end{figure}
\begin{figure}
\begin{center}
\vspace*{-0.6cm}
\epsfig{figure=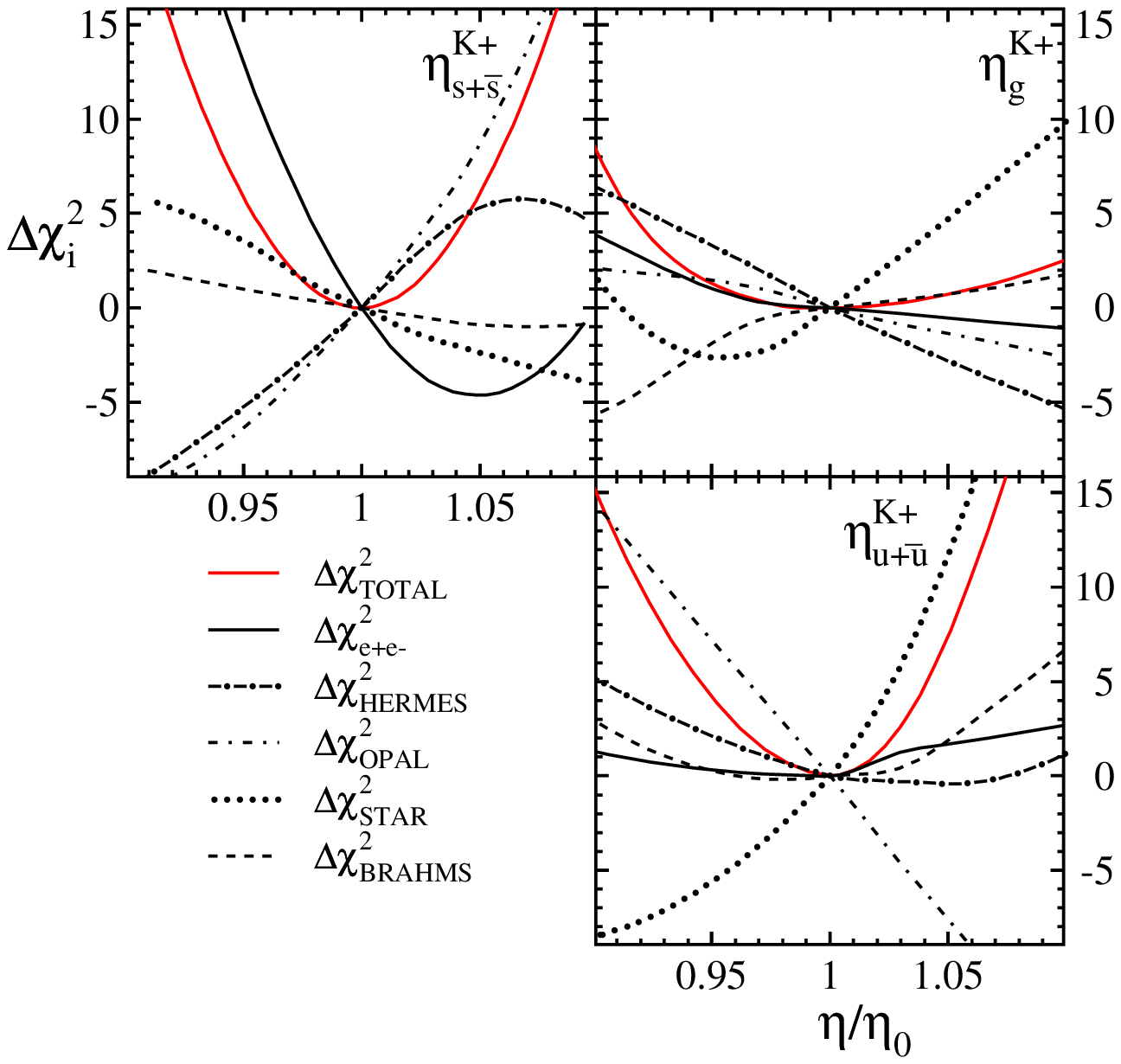,width=0.495\textwidth}
\end{center}
\vspace*{-0.7cm}
\caption{The same as in Fig.~\ref{fig:pion-partial}, but for NLO kaon fits.
\label{fig:kaon-partial}}
\vspace*{-0.5cm}
\end{figure}

In order to give a clear and comprehensive picture of the typical
uncertainties characteristic of the fragmentation functions obtained
in the global fits, in the present Section we apply the Lagrange
multiplier technique introduced in Sec.~\ref{sec:lagrange}.

Rather than focusing on the uncertainties of the parameters 
in Eq.~(\ref{eq:ff-input}) determining the fragmentation functions 
at the initial scale and choosing a particular increment $\Delta \chi^2$
to judge the quality of the fit, we find it much more enlightening to
analyze the range of variation of other relevant features of the
fragmentation functions, with a more apparent physical meaning, and
take these as the characteristic uncertainties of the fit.
Notice that the range of variation of the fitted parameters are strongly
correlated; the impact of any of them on the
behavior of the distributions, or on a given observable, is determined
also by the values taken by the whole set of parameters through the
evolution equations.
Of course, in order to get a precise
estimate of the uncertainty in a given observable computed with the
set, the range of variation of that particular observable as a function
of $\Delta \chi^2$ has to be evaluated. As explained in Sec.~\ref{sec:lagrange},
the result takes into account the complex correlations between the
parameters, implies no assumptions on the profile of $\chi^2$, and allows to
consider different $\Delta \chi^2$.

In Figures~\ref{fig:pion-uncert} and \ref{fig:kaon-uncert} we show,
as an example, the range of variation of the truncated second
moments of the fragmentation functions
\begin{equation}
\label{eq:truncmom}
\eta^H_i(x_p,Q^2) \equiv \int_{x_p}^1 z D_i^H(z,Q^2) dz, 
\end{equation}
for $x_p=0.2$ and $Q=5\, \text{GeV}$, around the values obtained
for them in the best fit to data, $\eta^H_{i\,0}$, against the
corresponding increase in $\chi^2$. In the lowest order, the second
moments represent the energy fraction of the parent parton of flavor $i$ 
taken by the hadron $H$. The truncated moments discard the
low-$z$ contributions, which are not constrained by data 
and therefore only determined by low-$z$ extrapolation.

As it can be seen in the upper left panel of Fig.~\ref{fig:pion-uncert},
the truncated moment $\eta^{\pi^+}_{u+\overline{u}}$,
associated with $D^{\pi^+}_{u+\overline{u}}$, is the constrained best,
with a range of variation of less than 3\% around
the value computed with the best fit, assuming a very conservative increase
in $\chi^2$ by 15 units, i.e., $\Delta \chi^2=15$.
This comparatively stringent restriction comes from
the fact that all the observables accounted for in the fit have a strong
dependence on the corresponding $D^{\pi^+}_{u+\overline{u}}$ fragmentation function. Moving to
the next panel, we find that for the unfavored  $D^{\pi^+}_{\overline{u}}$
fragmentation function, the truncated moment is less, but still well
constrained within a 5\% range for a similar $\Delta \chi^2$.
While this distribution cannot be determined by SIA data without flavor
symmetry assumptions, it is constrained by SIDIS and proton-proton collisions
involving low $x$ contributions, where the $\overline{u}(x)$ PDF is large, or at high
$x$ through its alter ego $D^{\pi^-}_{u}$.

The fragmentation functions for strange quarks have much larger uncertainties,
and their moment can vary by more that 10\%, as shown in the upper right panel of
Fig.~\ref{fig:pion-uncert}. In our fit, not only
$D^{\pi^+}_{\overline{u}} \neq D^{\pi^+}_{\overline{s}}$ but their
respective uncertainties are found to be quite different as well. Notice
that these uncertainties in the unfavored distributions cannot
justify the differences with KRE and AKK sets found in the previous
Section.

The moment for the gluon fragmentation function, in the lower left panel of
Fig.~\ref{fig:pion-uncert}, is restricted to vary by less than 10\%,
with the constraint mainly stemming from the evolution from the
initial scale to the scales relevant for each measurement, rather than from a direct
contribution to a particular cross section. Such kind of contributions
are certainly present in RHIC data, but only in narrow intervals of $z$, 
so they cannot fix the truncated moment for $x_p=0.2$ by themselves. For heavy quarks, flavor
tagged data dominate the fit, and the more precise b-tagged data
lead to better constrained moments for $D^{\pi^+}_{b+\overline{b}}$
than for $D^{\pi^+}_{c+\overline{c}}$.

The uncertainties in kaon fragmentation functions, Fig.~\ref{fig:kaon-uncert},
are typically twice as large as those for pions, with 10\% variations for the 
total fragmentation functions
containing a favored fragmentation function and rather poorly constrained
unfavored fragmentation functions
$D^{K^+}_{\overline{u}}=D^{K^+}_{\overline{d}}=D^{K^+}_{d}$. 
The profiles show much more significant deviations from the parabolic behavior 
than in the case of pions, and the upper bound for the
moment of the gluon distribution is much less defined.

Next we further illuminate the role of the different data sets
in setting the constraints on the fragmentation functions. 
In Figs.~\ref{fig:pion-partial} and \ref{fig:kaon-partial} we show the
partial contributions $\Delta \chi^2_i$ of a data set $i$ to the increase of 
$\chi^2$ from its minimum value defined by the best fit against the
variation of some selected truncated moments.

In the upper left panel of
Fig.~\ref{fig:pion-partial} it can be noticed that both SIA (thick
solid line) and SIDIS (thick dashed dotted line) data define their own
minima for the $u+\overline{u}$ moment, slightly to the
right and to the left of the best fit value, respectively, but within
the above mentioned conservative uncertainty range of $\Delta\chi^2=15$.
The other data sets fail to develop well defined minima (at least within the
shown range), and the final result is a compromise between all of them.

Notice that the failure to define a minimum by a given subset of data may
not necessarily imply a weak dependence of the data on the particular
flavor that fragments. It may happen because the data cover a limited range in $z$, while
the observable we chose picks up contributions over a much wider range in $z$.
The seemingly contradictory ``preferences'' of two data sets can follow from
sensitivity to complementary regions in $z$.

The truncated moment for the gluon-to-pion fragmentation function, 
shown in the upper right panel of Fig.~\ref{fig:pion-partial},  
happens to be well constrained again by
SIA and SIDIS data, with their respective minima very close to the best
fit result. For $\eta_{s+\overline{s}}^{\pi^+}$, in the lower
right panel, neither set shows a minimum and the final result is
a compromise. Notice that, here, the most significant
contributions are those coming from SIA and the OPAL tagging
probabilities.

Partial contributions $\Delta\chi^2_i$ to the truncated moments
for kaon fragmentation functions, Fig.~\ref{fig:kaon-partial}, have less definite preferences, 
leading to much larger uncertainties. Starting with the moment for $s+\overline{s}$
in the upper left panel, we find that SIDIS data now fail to define a minimum.
This is mainly due to the limited strangeness content in the proton. However, 
SIDIS contributes with OPAL tagged data to balance the preference coming from
SIA data, which is slightly larger than the best fit value. Gluon-to-kaon fragmentation, 
shown in the upper right panel, is mostly constrained by STAR and SIDIS data.
The $u+\overline{u}$ moment, receiving contributions from the suppressed
$D^{K^+}_u$ and the doubly suppressed $D^{K^+}_{\overline{u}}$, is neither well
constrained by SIA nor by SIDIS data, however, OPAL and
STAR data help to improve the situation.

\begin{figure}[ht!]
\begin{center}
\vspace*{-0.6cm}
\epsfig{figure=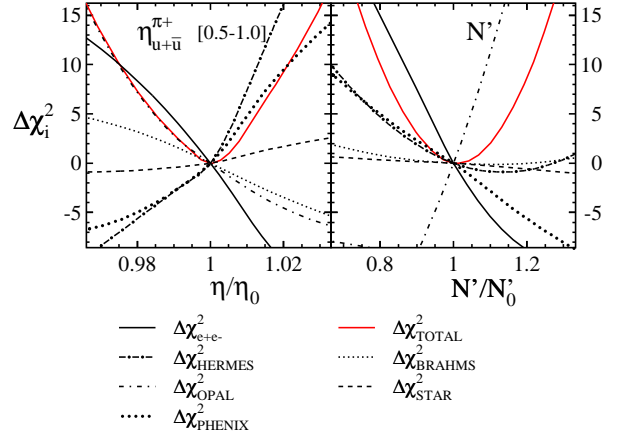,width=0.495\textwidth}
\end{center}
\vspace*{-2.5cm}
\caption{The same as in Fig.~\ref{fig:pion-partial}, but for
$\eta^{\pi^+}_{u+\overline{u}}(x_p=0.5,Q=5\,\text{GeV})$ (left panel) and
$N^{\prime}$ (right panel).
\label{fig:pion-partialz}}
\vspace*{-0.5cm}
\end{figure}

Upon the completion of our analysis, a new determination of fragmentation
functions, again based only on SIA data but with a careful assessment 
of uncertainties, was published \cite{ref:hirai}. There, it is shown that the large
differences found between the most widely used sets (KRE, KKP, and AKK)
are related to the large $z$ behavior of the fragmentation functions, where the
uncertainties are found to be most significant. Indeed, we can see that this is the
case for sets based only in SIA data, but the situation 
is considerably improved in a global fit. In the left panel of Fig.~\ref{fig:pion-partialz}
we show, as an example, the partial contributions $\Delta\chi^2_i$
from different sets of data to the truncated moment for $D^{\pi^+}_{u+\overline{u}}$, 
but now taking only large-$z$ contributions, $x_p=0.5$ in Eq.~(\ref{eq:truncmom}). 
Here it can be seen that SIA data, as expected, 
fail to define a minimum, but the complementary information coming from
the other sets, dominated by PHENIX and SIDIS, define a clear minimum,
with a well a constrained range of variation, again showing the power
and importance of a global analysis.

Another interesting difference with the analysis of
Ref.~\cite{ref:hirai} is that while they accomplish a good fit to
SIA data under the {\em assumption} $N^{\prime}=1$ in Eq.~(\ref{eq:sea_break}), our 
global fit prefers $N^{\prime}\simeq 0.83$. In order to understand 
this difference, the right panel of Fig.~\ref{fig:pion-partialz} shows the partial
contribution to $\chi^2$ for different values of $N^{\prime}$
normalized to $N^{\prime}$ for our best fit. 
Although the fit clearly prefers $N^{\prime}\simeq 0.83$,
the uncertainty is large, and for a very conservative $\Delta
\chi^2$ may be taken as marginally consistent with $N^{\prime}=1$.
The preference for $N^{\prime}$ values smaller than one in our global
fit is driven by the OPAL tagging probabilities, not included in
\cite{ref:hirai}.

Finally, we note that projected measurements of hadron production in SIA by 
the BELLE and/or the BaBar experiments at low c.m.s.\ energies would open up 
the possibility for
studies of scaling violations with unprecedented precision. Such data
should help to further constrain the fragmentation functions and significantly
reduce the present uncertainties.

\section{Conclusions}

We have demonstrated the feasibility of a NLO combined QCD analysis of single
inclusive hadron production data for pions and kaons, 
coming from electron-positron annihilation, deep-inelastic lepton-nucleon scattering, 
and proton-proton collisions, collected over a wide kinematic range.

At variance with previous fits based only on
electron-positron annihilation data, the present analysis includes
complementary information from other experiments that reduce
significantly the uncertainties of the resulting fragmentation functions.

In the case of pion fragmentation functions, we find that the new
SIDIS data provided by the HERMES experiment, effectively constrain the separation
between favored and unfavored distributions, a separation that was either
not implemented in previous sets or it was based on certain assumptions.
The most recent RHIC results provide stringent constraints on the
gluon fragmentation function and, in general, on the large $z$ behavior
of the other distributions.
For kaons, the new data modify significantly the up-to-now standard
picture provided by previous analyses. Specifically, SIDIS data rule out
the flavor separation scheme hitherto implemented, while RHIC data lead to
a new gluon fragmentation function, and thus scale dependences, significantly
different over the whole range of $z$.

The implementation of the $\chi^2$ minimization in our global analysis is
numerically fast and efficient and can be straightforwardly expanded to any
future set of hadron production data. With the help of the Mellin moment technique,
the entire analysis was consistently performed at NLO accuracy without
resorting to often used approximations for NLO hard scattering cross sections.

The success of the global analysis performed here, for the first time
including observables other than single inclusive annihilation, stands
for an explicit check of factorization, universality, and the 
perturbative QCD framework for the description of the corresponding processes, 
providing at the same time much more precise constraints on the fragmentation functions.
Proton-proton collision data and that coming from SIDIS offer a
crucial piece of information that cannot be disregarded and will be
increasingly accessible in the near future.

\acknowledgments
We warmly acknowledge Elke Aschenauer and Achim Hillenbrand for 
help with the HERMES data, Werner Vogelsang for helpful discussions, 
and Carlos Garc\'{\i}a Canal for comments and suggestions.
This work was partially supported by CONICET, ANPCyT and UBACyT.



\begin{thebibliography}{99}
%
\bibitem{ref:framework} See, e.g., J.C.\ Collins, D.E.\ Soper, and G.\ Sterman, 
``Perturbative QCD'', A.H.\ Mueller (ed.),
Adv. Ser. Direct. High Energy Phys. {\bf 5}, 1 (1988) and references therein.
%
\bibitem{ref:collins-soper} J.C.\ Collins and D.E.\ Soper, Nucl. Phys.
{\bf B193}, 381 (1981); {\bf B213}, 545(E) (1983); {\bf B194}, 445 (1992).
%
\bibitem{Campbell:2006wx}
  J.M.~Campbell, J.W.~Huston, and W.J.~Stirling,
  Rept.\ Prog.\ Phys.\  {\bf 70}, 89 (2007).
%
\bibitem{Vogt:2006bt} See, e.g., A.~Vogt, S.~Moch, and J.~Vermaseren,
  Nucl.\ Phys.\ Proc.\ Suppl.\  {\bf 160}, 44 (2006),
%
\bibitem{Thorne:2006zu}
  R.S.~Thorne, A.D.~Martin, and W.J.~Stirling,
  {\tt hep-ph/0606244}.
%
\bibitem{Pumplin:2005rh}
  J.~Pumplin, A.~Belyaev, J.~Huston, D.~Stump, and W.-K.~Tung,
  JHEP {\bf 0602}, 032 (2006).
%
\bibitem{ref:kretzer} S.\ Kretzer, Phys. Rev. {\bf D62}, 054001 (2000).
%
\bibitem{ref:kkp} B.A.\ Kniehl, G.\ Kramer, and B.\ P\"{o}tter,
Nucl. Phys. {\bf B582}, 514 (2000).
%
\bibitem{ref:akk} S.\ Albino, B.A.\ Kniehl, and G.\ Kramer,
Nucl. Phys. {\bf B725}, 181 (2005); {\bf B734},50 (2006).
%
\bibitem{ref:hirai} M.\ Hirai, S.\ Kumano, T.-H.\ Nagai, and
K.\ Sudoh, {\tt hep-ph/0702250}.
%
\bibitem{ref:alephdata} ALEPH Collaboration, D.\ Buskulic {\em et al.},
Z. Phys. {\bf C66}, 355 (1995).
%
\bibitem{ref:delphidata} DELPHI Collaboration, P.\ Abreu {\em et al.},
Eur. Phys. J. {\bf C5}, 585 (1998).
%
\bibitem{ref:opaldata} OPAL Collaboration, R.\ Akers {\em et al.},
Zeit. Phys. {\bf C63}, 181 (1994).
%
\bibitem{ref:opaleta} OPAL Collaboration, G.\ Abbiendi {\em et al.},
Eur. Phys. J. {\bf C16}, 407 (2000).
%
\bibitem{ref:tpcdata} TP Collaboration, H.\ Aihara {\em et al.},
Phys. Rev. Lett. {\bf 61}, 1263 (1998); Phys. Lett. {\bf B184}, 299
(1987); X.-Q.\ Lu, Ph.D.\ thesis, John Hopkins University,
UMI-87-07273, 1986.
%
\bibitem{ref:slddata} SLD Collaboration, K.\ Abe {\em et al.},
Phys. Rev. {\bf D59}, 052001 (1999).
%
\bibitem{ref:tassodata} TASSO Collaboration, W.\ Braunschweig {\em et al.},
Zeit. Phys. {\bf C42}, 189 (1989).
%
\bibitem{ref:hermessidis}   A.\ Hillenbrand (HERMES Collaboration),
``Measurement and Simulation of the Fragmentation Process at HERMES'',
Ph.D.\ thesis, Erlangen Univ., Germany, September 2005;
private communications.
%
\bibitem{ref:phenixpion} PHENIX Collaboration, S.S.\ Adler {\em et al.},
Phys. Rev. Lett. {\bf 91}, 241803 (2003); K. Barish (PHENIX Collaboration),
``The PHENIX Spin Program: Recent Results and Future Prospects'',
talk presented at SPIN 2006, Kyoto, Japan. 
%
\bibitem{ref:starpion} STAR Collaboration, J.\ Adams {\em et al.},
Phys. Rev. Lett. {\bf 97}, 152302 (2006).
%
\bibitem{ref:brahms} BRAHMS Collaboration, I.\ Arsene {\em et al.},
{\tt hep-ex/0701041}.
%
\bibitem{ref:starkaon} STAR Collaboration, B.I.\ Abelev {\em et al.},.
{\tt nucl-ex/0607033}.
%
\bibitem{ref:Stump:2001gu} J.~Pumplin, D.R.~Stump, and W.-K.~Tung,
  Phys.\ Rev.\ {\bf D65}, 014011 (2002); D.~Stump {\it et al.},
  Phys.\ Rev.\ {\bf D65}, 014012 (2002).
%
\bibitem{ref:dsv-lambda} D.\ de Florian, M.\ Stratmann, and W.\ Vogelsang,
Phys. Rev. {\bf D57}, 5811 (1998).
%
\bibitem{ref:jsv-pion} B.\ J\"{a}ger, A.\ Sch\"{a}fer, M.\ Stratmann, and
W.\ Vogelsang, Phys. Rev. {\bf D67}, 054005 (2003).
%
\bibitem{ref:rijken} P.J.\ Rijken and W.L.\ van Neerven, Nucl. Phys.
{\bf B487}, 233 (1997); Phys. Lett. {\bf B386}, 422 (1996).
%
\bibitem{ref:nlo-kernels} G.\ Curci, W.\ Furmanski, and R.\ Petronzio, Nucl. Phys.
{\bf{B175}}, 27 (1980);
W.\ Furmanski and R.\ Petronzio, Phys. Lett. {\bf{97B}}, 437 (1980);
L.\ Beaulieu, E.G.\ Floratos, and C.\ Kounnas, Nucl. Phys. {\bf B166}, 321 (1980).
%
\bibitem{ref:sv-kernels} M.\ Stratmann and W.\ Vogelsang,
Nucl. Phys. {\bf B496}, 41 (1997).
%
\bibitem{ref:aempi} G.\ Altarelli, R.K.\ Ellis, G.\ Martinelli, and S.Y. Pi,
Nucl. Phys. {\bf B160}, 301 (1979);
P.\ Nason and B.\ Webber, Nucl. Phys. {\bf B421}, 473 (1994);
{\bf B480}, 755(E) (1996).
%
\bibitem{ref:fupe} W.\ Furmanski and R.\ Petronzio, Z. Phys. {\bf C11}, 293 (1982).

%
\bibitem{ref:graudenz} D.\ Graudenz, Nucl. Phys. {\bf B432}, 351 (1994).
%
\bibitem{ref:targetfrag} L.\ Trentadue and G.\ Veneziano,
Phys. Lett. {\bf B323}, 201 (1993);
D.\ de Florian, C.\ Garc\'\i a Canal, and R.\ Sassot,
Nucl. Phys. {\bf B470}, 195 (1996); D.\ de Florian and R.\ Sassot,
Nucl. Phys. {\bf B488}, 367 (1997).
%
\bibitem{ref:aversa} F.\ Aversa, P.\ Chiappetta, M.\ Greco, and J.-Ph.\ Guillet,
Nucl. Phys. {\bf B327}, 105 (1989);
D.\ de Florian, Phys. Rev. {\bf D67}, 054004 (2003).
%
\bibitem{ref:kretzer2} E.\ Christova, S.\ Kretzer, and E.\ Leader,
Eur. Phys. J. {\bf C22}, 269 (2001).
%
\bibitem{ref:cacciari} M.\ Cacciari, P.\ Nason, and C.\ Oleari,
JHEP {\bf 0510}, 034 (2005).
\bibitem{ref:mrst} A.D.\ Martin, R.G.\ Roberts, W.J.\ Stirling, and R.S.\ Thorne,
Eur. Phys. J. {\bf C28}, 455 (2003).
%
\bibitem{ref:mrstlo} A.D.\ Martin, R.G.\ Roberts, W.J.\ Stirling, and R.S.\ Thorne,
Phys. Lett. {\bf B531}, 216 (2002).
\bibitem{ref:whalley} For a compilation of SIA data, see,
G.D.\ Lafferty, P.I.\ Reeves, and M.R.\ Whalley, J. Phys. {\bf G21}, A1 (1995).
%
\bibitem{ref:grv1} M.\ Gl\"{u}ck, E.\ Reya, and A.\ Vogt,
Z. Phys. {\bf C48}, 471 (1990); Phys. Rev. {\bf D45}, 3986 (1992).
%
\bibitem{ref:grv2} M.\ Gl\"{u}ck, E.\ Reya, and A.\ Vogt,
Phys. Rev. {\bf D48}, 116 (1993); {\bf D51}, 1427(E) (1995).
%
\bibitem{ref:mellin} M.\ Stratmann and W.\ Vogelsang, Phys. Rev. {\bf D64},
114007 (2001).
%
\bibitem{ref:mellin2} C.\ Berger, D.\ Graudenz, M.\ Hampel, and A.\ Vogt,
Z. Phys. {\bf C70}, 77 (1996); D.A.\ Kosower, Nucl. Phys. {\bf B520} (1998) 263.
%
\bibitem{ref:BOTJE}M. Botje, Eur. Phys. J. {\bf C14} (2000) 285.
%
\bibitem{ref:polpdf} D.~de Florian, G.A.~Navarro, and R.~Sassot,
  Phys.\ Rev.\  {\bf D71}, 094018 (2005);
  G.A.~Navarro and R.~Sassot, Phys.\ Rev.\  {\bf D74}, 011502 (2006);
  X.~Jiang, G.A.~Navarro, and R.~Sassot, Eur.\ Phys.\ J.\  {\bf C47}, 81 (2006).
%
\bibitem{ref:guzey} V.\ Guzey, M.\ Strikman, and W.\ Vogelsang, Phys. Lett. {\bf B603}, 173 (2004).
%
\bibitem{Aktas:2004rb} H1 Collaboration, A.~Aktas {\it et al.},
Eur.\ Phys.\ J.\  {\bf C36}, 441 (2004).
%
\bibitem{Daleo:2004pn} A.~Daleo, D.~de Florian, and R.~Sassot,
  Phys.\ Rev.\ {\bf D71}, 034013 (2005).
\end{thebibliography}
\end{document}